\documentclass[a4paper,11pt]{article}

\pdfoutput=1
\usepackage{jheppub}
\usepackage[T1]{fontenc}
\usepackage{graphicx,amstext,amsmath,amssymb,epsfig,color,ulem,color}

\newcommand{\gsim}{\lower.7ex\hbox{$\;\stackrel{\textstyle>}{\sim}\;$}}
\newcommand{\lsim}{\lower.7ex\hbox{$\;\stackrel{\textstyle<}{\sim}\;$}}

\newcommand{\eqn}[1]{Eq.~(\ref{#1})}
\newcommand{\fig}[1]{Fig.~(\ref{#1})}

\def\be{\begin{equation}} 
\def\ee{\end{equation}} 
\def\del{\partial} 
\def\N{\mathcal{N}}
\def\Cc{\mathcal{C}}
\def\lGB{\lambda_{GB}}

\usepackage{mathtools}
\usepackage{hyperref} 
\usepackage{soul}
\hypersetup{
    unicode=true,          
    bookmarksnumbered=true,
    colorlinks=true,       
    linkcolor=red,          
    citecolor=[rgb]{0,.7,0}, 
    filecolor=magenta,
    urlcolor=cyan,
    menucolor=blue,
    baseurl=http://www.arxiv.org/abs/,
    linktocpage=true,
    hyperfootnotes=true
}

\title{\boldmath Resurgence and Hydrodynamic Attractors in Gauss-Bonnet Holography}

\author{Jorge Casalderrey-Solana,}
\author{Nikola I. Gushterov,}
\author{Ben Meiring}

\affiliation{Rudolf Peierls Centre for Theoretical Physics, University of Oxford,\\1 Keble Road, Oxford OX1 3NP, United Kingdom}

\emailAdd{jorge.casalderreysolana@physics.ox.ac.uk}
\emailAdd{nikola.gushterov@balliol.ox.ac.uk}
\emailAdd{ben.meiring@physics.ox.ac.uk}

\abstract{
We study the convergence of the hydrodynamic series in the gravity dual of Gauss-Bonnet gravity in five dimensions with negative cosmological constant via holography. By imposing boost invariance symmetry, we find a solution to the Gauss-Bonnet equation of motion in inverse powers of the proper time, from which we can extract high order corrections to Bjorken flow for different values of the Gauss-Bonnet parameter $\lGB$.
As in all other known examples the gradient expansion is, at most, an asymptotic series which can be understood through applying the techniques of Borel-Pad\'e summation.
As expected from the behaviour of the quasi-normal modes in the theory, we observe that the singularities in the Borel plane of this series show qualitative features that interpolate between the infinitely strong coupling limit of $\N=4$ Super Yang Mills theory and the expectation from kinetic theory.  We further perform the Borel resummation  to constrain the behaviour of hydrodynamic attractors beyond leading order in the hydrodynamic expansion. We find that for all values of  $\lGB$  considered, the convergence of different initial conditions to the resummation and its hydrodynamization occur at large and comparable values of the pressure anisotropy.
}

\begin{document} 
\maketitle
\flushbottom


\section{Introduction}
\label{Intro}
In recent years there has been increasing interest in understanding the emergence of hydrodynamic behaviour in relativistic theories. 
In addition to many new theoretical advances, the strong multi-particle correlations observed in heavy ion collisions both at RHIC \cite{Ackermann:2000tr,Adler:2003kt,Back:2004mh} and the LHC \cite{ATLAS:2012at,Chatrchyan:2012ta,Aamodt:2010pa} and its successful description via hydrodynamical modelling 
\cite{Huovinen:2001cy,Teaney:2001av,Hirano:2005xf,Schenke:2010rr,Hirano:2010je,Shen:2014vra}
has provided a testing ground to explore how this collective behaviour arises from a microscopic theory. One of the most surprising empirical insights 
that this type of modelling 
of subnuclear dynamics has revealed is the fact that hydrodynamics can describe the bulk properties of the system even for extreme pressure gradients, which at face value, question the applicability of this long distance effective theory to those collisions. The recent observation of collective phenomena also in even smaller systems, such as  p-p collisions \cite{Khachatryan:2010gv,Aad:2015gqa,Khachatryan:2016txc} together with the success of the same hydrodynamic modelling in describing them \cite{Bozek:2011if,Kozlov:2014fqa,Weller:2017tsr} poses new challenges to our understanding of the applicability of this theory.

In conjunction with the above phenomenological observations, 
in recent years  several numerical experiments have been performed to test the validity of hydrodynamics in different ultra-reltavisitic scenarios.
Both in the infinitely strongly coupled limit of $\N=4$ SYM, described via holography, \cite{Heller:2011ju,Chesler:2009cy,Chesler:2010bi,Chesler:2015wra} and in the weakly coupled (perturbative) limit of gauge theories, described via kinetic theory \cite{Kurkela:2015qoa}, the direct comparison of the full stress tensor in different out of equilibrium processes with the 
hydrodynamic expectation showed that hydrodynamics can provide an accurate description of the evolution of the system even where the gradient terms are as large as the leading order terms. These experiments demonstrate that hydrodynamics can be applied even if the system under consideration is very far away from local thermal equilibrium and with strong deviations from the equation of state, as explicitly demonstrated in the analysis of non-conformal theories \cite{Attems:2016tby} .

Complementary  to these numerical studies, the convergence of the hydrodynamic series has been recently analysed. In a seminal paper \cite{Heller:2013fn}, the late time behaviour of boost invariant expansion of $\N=4$ SYM was analysed in a power series of the inverse proper time up to very high order. This series may be viewed as an expansion in the Knudsen number, and the coefficients of the series are controlled by increasing orders in the hydrodynamic expansion. The analysis of these large order perturbations showed that the hydrodynamic expansion is an asymptotic series, exhibiting a factorial growth of the series coefficients. Similar behaviour was found in different ultraviolet completions of second order hydrodynamics \cite{Heller:2015dha,Basar:2015ava,Aniceto:2015mto} and  kinetic theory in the relaxation time approximation (RTA) \cite{Heller:2016rtz}  (see also \cite{Denicol:2016bjh} for a complementary analysis of the convergence of the hydrodynamic series).  
Interestingly, the analysis of these series via Borel-Pad\'e techniques showed that these large order gradient expansions are sensitive to non-hydrodynamic modes, which play an equivalent role to non-perturbative corrections in perturbation theory. 

Numerical analyses of these same boost invariant flows led Heller and Spali{\'n}ski to suggest the existence of a hydrodynamic attractor  \cite{Heller:2015dha} which may be thought as an extension of hydrodynamics beyond local thermal equilibrium \cite{Romatschke:2017vte}. These are special time dependent configurations to which all other boost invariant evolutions of the system converge at different times . These solutions have been found in different theories, such as 
Israel-Stewart (IS) \cite{Israel:1979wp} and Baier-Romatschke-Son-Starinets-Stephanov (BRSSS) \cite{Baier:2007ix} 
 hydrodynamics \cite{Heller:2015dha,Denicol:2017lxn}, $\N=4$ SYM, kinetic theory \cite{Romatschke:2017vte} or anisotropic hydrodynamics \cite{Strickland:2017kux}. It has been further argued that these special solutions may be found, or at least well approximated, via a trans-series solution of the system evolution \cite{Heller:2015dha,Spalinski:2017mel}, that non-perturbatively completes the gradient expansion.
 These types of solutions have also been recently analysed in less symmetric situations, including non-conformal theories and less symmetric flows \cite{Romatschke:2017acs,Behtash:2017wqg}.  Quite remarkably in all those cases the hydrodynamic attractor is very well approximated up to unexpectedly large values of the gradient by first order hydrodynamics, providing a dynamical understanding to the unexpected success of this theory. 

The emergence of these solutions has only been studied in the two extreme cases of infinitely strong and perturbatively weak coupling. However, to better connect these theoretical advances with phenomenological applications it is important to understand how these special types of solutions behave at intermediate coupling. Starting from the infinite coupling limit, 
finite coupling corrections are studied  via the gauge/gravity duality by introducing higher curvature terms in the dual gravitational theory \cite{Gubser:1998nz,Pawelczyk:1998pb}. For the gravity dual of $\N=4$ SYM, the relevant correction that affects the dynamics of the field theory stress tensor are expressed in terms of the Weyl tensor and are quartic in the curvature. This implies that those higher derivative terms can only be studied perturbatively, to avoid the emergence of instabilities, ghosts and other pathologies. Within this limit,
these types of corrections have been vigorously studied in the past, exploring the correction of many different quantities (see \cite{Waeber:2015oka} for a recent compilation of results). Recently, the relaxation of small stress tensor fluctuations in the thermal ensemble of $\N=4 $ SYM has been studied in detail \cite{Grozdanov:2016vgg,Grozdanov:2016fkt}, (see also \cite{Stricker:2013lma,Steineder:2013ana} for previous studies).  In these studies, a new set of purely dissipative modes have been found, which are an intrinsic consequence of finite coupling. 

Another higher-derivative theory which has also received significant attention in the holographic context is Gauss-Bonnet gravity. The action of this theory includes both quadratic and quartic curvature terms, governed by a single parameter 
$\lGB$; 
nevertheless, as a Lovelock theory, its classical equations of motion contain only up to quadratic derivatives of the metric, which in principle allows the non-perturbative treatment of the high-derivative terms. Unfortunately, the dual field theory to Gauss-Bonnet gravity is unknown. Nevertheless, in this holographic construction it is possible to extract static (thermodynamic) and transport properties of the putative dual field theory, as well as the relaxation of small non-thermal perturbations \cite{Grozdanov:2016vgg,Grozdanov:2016fkt},  its off-equilibrium dynamics \cite{Grozdanov:2016zjj,Andrade:2016rln} and corrections to its hydrodynamic expansion \cite{DiNunno:2017obv}. The comparison of those analysis and the results obtained from finite coupling corrections of $\N=4$ SYM show that for negative $\lGB$ these two theories share many common qualitative aspects, which makes Gauss-Bonnet holography an interesting laboratory with which to explore finite coupling effects to holographic theories. Note however that causality, positivity of energy and hyperbolicity considerations constrain the range of values of $\lGB$
\cite{Buchel:2008vz,Hofman:2008ar,Hofman:2009ug,Buchel:2009tt, Camanho:2014apa,Papallo:2015rna,Andrade:2016yzc,Konoplya:2017zwo}
 (see \cite{Grozdanov:2016fkt} for a detailed discussion on these limitations).

As a first step towards intermediate coupling, in this paper we analyse the convergence of the hydrodynamic series in the gradient expansion
 of matter dual to Gauss-Bonnet holography. As in the previously mentioned examples, we find that the series is asymptotic as a consequence of non-perturbative (in gradient strength) contributions, given by the quasi-normal modes of the dual black-hole. Remarkably, these characteristic modes dictate that the structure of singularities of the Borel transform associated to the hydrodynamic expansion interpolate between the known examples of infinitely strongly and weakly coupled theories. After characterising the analytic properties of the Borel transform, we resum the hydrodynamic series and constrain the dynamics of the hydrodynamic attractor. 
For all values of $\lGB$ considered, the resummation approaches first order hydrodynamics even when gradient corrections to the stress tensor are large and at comparable values of the pressure anisotropy. Therefore, even though as the viscosity increases the approach to hydrodynamics occurs at a decreasing value of the gradient, this hydrodynamization processes takes place at comparable values of a viscosity-rescaled gradient, as suggested in  \cite{Keegan:2015avk}. Analysing the leading contributions in the trans-series, we estimate how close the resummation is to the attractor solution. 
By varying the amplitude of those non-perturbative corrections we observe that the convergence of different initial configurations towards the attractor occurs at values of the viscosity-rescaled gradient for which the 
resummation has hydrodynamized. We also perform an identical analysis of the hydrodynamic series of RTA kinetic theory \cite{Heller:2016rtz} and find that in this model both hydrodynamization and convergence to the attractor solution occurs at smaller viscosity-rescaled gradients than in the strong coupling computations.

This paper is organised as follows: in Section~(\ref{sec:review}) we review the main properties of boost invariant flow and holographic Gauss-Bonnet to set up the notation, define the main quantities and outline the strategy to compute large order gradient corrections in this higher-curvature theory.  In Section~(\ref{sec:resurgence}) we perform a Borel transformation of the hydro expansion and analyse the phenomenon of resurgence, {\it i.e.} the emergence of non-perturbative information in the perturbative series,  in this context. After analytically continuing the Borel series, in 
Section~(\ref{sec:resum}) we resum the hydrodynamic expansion and constrain the dynamics of the attractor by analysing the leading order terms in the trans-series. Finally,
 in Section~(\ref{sec:conclusions}) we put our results into context and conclude. 


\section{\label{sec:review} Boost Invariant Flow in Holographic Gauss-Bonnet Gravity}
\subsection{Boost Invariant Hydrodynamics}
Boost invariant flow, also known as Bjorken flow, is a particular class of solutions of ultra-relativistic hydrodynamics obtained by assuming that the dynamics of the fluid in 3+1 dimensions is independent of boost transformation in one of the space directions, $z$.  This solution was first described by Bjorken \cite{Bjorken:1982qr} as a model for particle production in high-energy hadronic collisions. This solution was motivated by the observation that particle productions in this type of collision exhibits relatively flat momentum-rapidity distributions; while deviations of this independence are observed, boost invariant solutions of hydrodynamics are routinely used in the phenomenology of heavy ion collisions (see \cite{Jeon:2015dfa} for recent review).

Hydrodynamics may be viewed as an approximation of the stress tensor in a gradient expansion around the local thermal equilibrium at each point, characterised by the local energy density  $\epsilon$, and the fluid velocity $U^{\mu}$ ,
\be
T^{\mu \nu}= \left(\epsilon + p(\epsilon) \right) U^\mu U^\nu + \eta^{\mu \nu} p(\epsilon) + \Pi^{\mu\nu} \, ,
\ee
where $p(\epsilon)$ is the equation of state. For conformal theories, such as the ones we will describe in this paper, this is given by 
$p(\epsilon)=\frac{1}{3} \epsilon$. The tensor $\Pi^{\mu\nu}$ encodes all the deviations of the stress tensor in a given state from local thermal equilibrium. In hydrodynamics, this tensor is expanded in space gradients\footnote{Space gradients are defined as the projection of the space-time gradient into the space components in the fluid rest fame, $\nabla^\mu=\Delta^{\mu\nu} \del_\nu$, where $\Delta^{\mu\nu}=\eta^{\mu\nu}-U^\mu U^\nu$.}
of the hydrodynamic fields as 
\be
\Pi^{\mu\nu}= -\eta \sigma^{\mu\nu} - \zeta \nabla^{\mu} U_\nu \eta^{\mu \nu} + ... \,, 
\ee
where $\eta$ and $\zeta$ are the first order transport coefficients, known as shear and bulk viscosities, and $\sigma^{\mu\nu}$ is the shear tensor, constructed from the symmetrised and traceless space gradient  of the velocity field.  In this expression the ellipses indicate additional gradient corrections which, as at first order,  can be expressed as a set of (independent) derivatives of the hydrodynamic fields multiplied by a 
set of transport coefficients which depend only on the local energy density. The complete set of second order coefficients and their operators can be found in \cite{Baier:2007ix,Bhattacharyya:2008jc} and \cite{Romatschke:2009kr} for conformal and non-conformal theories respectively.

Following Bjorken, a solution to leading order hydrodynamics equation ($\Pi^{\mu\nu}=0$) can be easily found assuming that the hydrodynamic fields depend only on one spatial direction, $z$, and that the fluid is boost invariant. Under this assumption, independently of the microscopic symmetries of the theory, entropy conservation of ideal hydrodynamics imposes that the entropy density behaves as  $s_0(\tau)\propto 1/\tau$, with  $\tau=\sqrt{t^2-z^2}$ the proper time. Focusing on conformal theories further simplifications can be made; in particular, dimensional analysis dictates that the local temperature and energy density are given respectively by 
$T(\tau)\propto \tau^{-1/3}$, $\epsilon \propto \tau^{-4/3}$. 
Similarly the size of the gradients can be expressed in terms the temperature $T(\tau)$, the only scale that 
characterises the state at a given proper time $\tau$. Therefore, the hydro expansion can be expressed as series of negative fractional powers of the proper time as 
\begin{equation} 
\label{bjorken}
\epsilon(\tau) = u^2  \Lambda^4 g_*\left(1 + \epsilon_{1} u +  \epsilon_{2} u^2 + ... \right) \,,
\end{equation}
with $u = \left(\Lambda \tau\right)^{-2/3}$, $\Lambda$ is a characteristic scale of mass dimension one, which in the hydrodynamic limit encodes all the information on the initial conditions of the flow and $\epsilon_i$ are dimensionless constants which are solely dependent on the degrees of freedom and transport coefficients, and are therefore a property of the theory, as opposed to the state. 
In particular, the constant we may choose $g_*=\epsilon_{\rm eq}/T^4$
 with $\epsilon_{\rm eq}$ the equilibrium energy density;  the leading gradient correction is controlled by the shear viscosity
and 
 similarly all additional coefficients are a combination of the transport coefficients up to the corresponding order.  In the absence of transverse dynamics, boost invariance further imposes that the velocity field of the fluid $U^\mu$ is given by $U^\tau=1$. 
Another remarkable aspect of this effectively 1-Dimensional flow is that once the functional form of the energy density is determined, stress tensor conservation and conformal symmetry completely fix the stress tensor of the theory. Defining the longitudinal ($P_L$) and transverse ($P_T$) pressures  as the diagonal components of the stress tensor in the fluid rest frame in the direction of expansion and perpendicular to it respectively, these are given by 
\be
P_L= -\epsilon (\tau) - \tau \dot \epsilon (\tau) \,, \quad P_T=\epsilon (\tau) + \frac{1}{2} \tau  \dot \epsilon (\tau) \,. 
\ee
Note that these expressions are valid independently of whether the system is far from local thermal equilibrium, and that in thermal equilibrium both pressures must be the same. Therefore, we may gauge how far the system is from local thermal equilibrium by monitoring how anisotropic the system becomes with respect to the average pressure, which we can determine by computing the anisotropy parameter
\be
\label{eq:Rdef}
R\equiv 3 \frac{P_T-P_L}{\epsilon} \,,
\ee
where the equilibrium pressure has been used to normalise the pressure anisotropy. 

The anisotropy parameter $R$ can be used to monitor the approach of a system to hydrodynamics \cite{Heller:2011ju}. 
Following the hydrodynamic expansion  \eqn{bjorken}, for a conformal theory the anisotropic parameter $R$ becomes a function of the dimensionless gradient,
\be
w= \tau T \,, 
\ee
where out of equilibrium we can identify the effective temperature as
\be
\label{eq:defTeff}
T(\tau) \equiv \left(\frac{\epsilon(\tau)}{g_*}\right)^{1/4}
\ee
where we have defined $g_*$ below  \eqn{bjorken}.
 In the hydrodynamic limit, $R$ admits an expansion in inverse powers of $w$ as
\be
\label{eq:Rseries}
R=\sum_{n=1} \frac{r_n}{w^n} \,, 
\ee
where, once again, the leading order term is controlled by the viscosity to entropy ratio $r_1=8\frac{\eta}{s}$ and, to first order in hydrodynamics, the anisotropy function is given by
\be
\label{eq:Rhyd}
R^{\rm 1st}_{\rm hyd}= 8 \frac{\eta}{s} \frac{1}{w}\,. 
\ee

Finally, for later reference, another dimensionless quantity which has been used in the literature for identifying attractor solution is the logarithmic proper time derivative of $w$
\be
\label{eq:fdef}
f\equiv \frac{\tau}{w} \frac{d}{d\tau} w= 1+ \frac{1}{4} \tau \frac{d}{d\tau} \log \epsilon \,.
\ee
The two function $f$ and $R$ are not independent of each other, since $R(w)=-12 + 18 f(w)$.

\subsection{Gauss-Bonnet Holography and Boost Invariant flow}

The Gauss-Bonnet gravity action is given by 
\begin{equation}
S =  \frac{1}{2 \kappa_5^2 } \int d^{5}x \,\sqrt{-g} \left(R + \frac{12}{L^2} + \frac{\lambda_{GB} L^2}{2}\left(R_{\mu\nu\rho\sigma}R^{\mu\nu\rho\sigma}-4 R_{\mu\nu}R^{\mu\nu}+ R^2\right)\right) \,,
\end{equation}
where $\kappa_5^2$ is proportional to the five dimensional Newton constant, $L$ is the AdS radius of the $\lGB=0$ theory, and $\lambda_{GB}$ is a dimensionless coupling which controls the magnitude of the higher derivative corrections. Without loss of generality, in the rest of this 
work we will set $L=1$.
 An important feature of this theory is that in spite of incorporating higher derivatives terms, black-brane solutions, dual to thermal ensembles of the associated gauge theories, can be found for non-perturbarive values of the Gauss-Bonnet parameter. From these the equation of state of the dual field theory can be extracted \cite{Gubser:1998nz,Pawelczyk:1998pb}
\begin{equation}
\label{eq:eos}
\epsilon = \frac{3}{8} \frac{\pi^2 N_c^2}{L_c^3} T^4 \,, \quad  L_c^2 = \frac{1+ \sqrt{1-4 \lambda_{GB}}}{2} \,. 
\end{equation}
As expected the $\lambda_{GB}\rightarrow 0$ limit of \eqn{eq:eos} agrees with the equation of state of $\mathcal{N}=4$ SYM. 
This expression indicates that not all values of $\lambda_{GB}$ are physical, since only 
$\lambda_{GB}\,  \in \left(-\infty,1/4\right]$ yield real energy densities. 
As already mentioned in the introduction, for arbitrary values of $\lGB$ this theory posseses causality problems associated with the superluminal propagation of high-momentum modes as well as negativity of the energy flux.  \cite{Brigante:2007nu,Buchel:2009tt}. These considerations
impose further constraints in the allowed values\footnote{
The analysis of three point functions of gravitons in high-derivatives theories has led the authors of \cite{Camanho:2014apa} to suggest that these theories are pathological for any strength of the higher derivative couplings unless a complete tower of stringy states is also considered. See however \cite{Papallo:2015rna} .
}
 of $-7/36< \lambda_{GB}\leq 9/100 $. 
 Nevertheless, since these constraints concern the ultraviolet behaviour of the theory, we may still consider  values of $\lGB$ beyond this region to explore its infrared dynamics, such as the approach towards hydrodynamics of the theory, as already done in \cite{Grozdanov:2016vgg,Grozdanov:2016fkt}. Note also that as in a strongly expanding system as boost invariant flows, the early time dynamics, $w\ll1$ are sensitive to these high-momentum pathological modes. For this reason in this work we will not explore the Bjorken flow dynamics at arbitrariy early times.

In addition to the equation of state, the transport coefficients of the holographic dual have also been analysed. In particular, the ratio of shear viscosity to entropy density is given by \cite{Brigante:2007nu}
\be
\frac{\eta}{s}=\frac{1-4 \lambda_{GB}}{4\pi} \,.
\ee
Second order transport coefficients of this theory have also been analysed in \cite{Grozdanov:2016fkt}. From this expression we can observe that positive values of $\lambda_{GB}$ yield smaller values of $\frac{\eta}{s}$ than $\N=4$ SYM \cite{Buchel:2008vz}. However, negative values of $\lambda_{GB}$ yield larger viscosity to entropy density ratios, as expected from finite t'Hooft coupling corrections of the infinite coupling limit in $\N=4$  SYM \cite{Buchel:2004di}. The analysis of the relaxation of small fluctuations of the thermal state via the computations of the quasi-normal mode spectrum of the dual black-branes indicates that many qualitative features of finite coupling corrections to $\N=4$ SYM are captured by Gauss-Bonnet holography with negative $\lGB$ \cite{Grozdanov:2016vgg}.
For this reason, in this paper we will only consider negative values of this parameter. 

Holographic duals of Bjorken-like flows in $\N=4$ have been explored by finding boost invariant solutions of the dual gravity theory \cite{Janik:2006gp,Benincasa:2007tp,Kinoshita:2008dq,Beuf:2009cx,Heller:2011ju}. This is achieved by imposing an Eddington-Finkelstein type ansatz  for the metric of the 5D space as  \cite{Kinoshita:2008dq}
\be
\label{eq:genBImetric}
 ds^2 = -r^2 A(\tau,r)d \tau^2 + 2 d r d \tau + (r \tau+1)^2 e^{b(\tau,r)} dy^2 + r^2 e^{ c(\tau,r)} d x_{\perp}^{2}
\ee
where $\tau=\sqrt{t^2-z^2}$ and $y={\rm arctanh} \left(z/t\right)$ are the standard proper time and rapidity coordinates and the asymptotically AdS boundary is located at $r\rightarrow \infty$.

Numerical solutions of the Einstein equations (with no higher derivative corrections, $\lGB=0$) with boost invariant symmetry from initial data at $\tau=0$ have been found in \cite{Heller:2011ju}. These solutions first showed the success of viscous hydrodynamics to describe the evolution of strongly coupled $\N=4$ SYM even when gradient corrections are large, later confirmed in less symmetric solutions. Boost invariant solutions of the Gauss-Bonnet equations of motion can also be found starting with this same ansatz. 
Imposing $AdS$ asymptotics (with radius $L_c$) leads to the following near boundary ($r\rightarrow \infty$) expansion of the different metric functions
\begin{align}
\label{eq:boundaryexpansion}
A(\tau,r) & =\frac{1}{L_{c}^{2}}+\frac{A^{(4)}(\tau)}{r^{4}}+...\\
b(\tau,r) & =-2\log\left(L_{c}\right)-\frac{2\left(1-L_{c}^{2}\right)}{r\tau}+\frac{1-L_{c}^{4}}{r^{2}\tau^{2}}-\frac{2\left(1-L_{c}^{6}\right)}{3r^{3}\tau^{3}}+\frac{b^{(4)}(\tau)}{r^{4}}+...\\
c(\tau,r) & =-2\log\left(L_{c}\right)+\frac{c^{(4)}(\tau)}{r^{4}}+...
\end{align}
 where the boundary value of the functions $b$ and $c$ are chosen such that the metric has AdS asymptotics  with radius $L_c$. The functions $A^{(4)}(\tau)$, $b^{(4)}(\tau)$
$c^{(4)}(\tau)$ cannot be determined from the near boundary expansion and additional infrared conditions, such as regularity, must be imposed. However, these are not all independent, since the power series expansion imposes
\begin{align}
b^{(4)}(\tau)=\frac{1-L_{c}^{8}}{2\tau^{4}} -2c^{(4)}(\tau) \,.
\end{align}
Energy-momentum conservation, which emerges from the boundary expansion as well, relates these functions to $A^{(4)}(\tau)$ which  must be extracted from the numerical computation.
 
From these solutions the dual field theory stress tensor can be extracted after holographic renormalisation, which has been performed for Gauss-Bonnet gravity in \cite{Brihaye:2008xu,Astefanesei:2008wz}. In terms of those functions, the stress tensor is diagonal and with the conventions $T_{ab}=\text{diag} \left(T_{\tau \tau}, \, T_{y y},\, T_{{\bf x}_\perp {\bf x}_\perp} \right)$ is given by 
\begin{equation}
\label{eq:Tabgen}
T_{ab} = \frac{N_c ^2}{2 \pi^2} \frac{(2L_{c}^{2}-1)}{L_{c}^{3}}\text{diag}\left(-\frac{3A^{(4)}}{4};~\tau^2\left(-\frac{2c^{(4)}}{L_{c}^{2}}-\frac{A^{(4)}}{4}\right);~~\frac{c^{(4)}}{L_{c}^{2}}-\frac{A^{(4)}}{4};~~\frac{c^{(4)}}{L_{c}^{2}}-\frac{A^{(4)}}{4}\right).
\end{equation}
This same ansatz has been used to obtain  the holographic equivalent of a gradient expansion \cite{Heller:2013fn}. Motivated by the expansion of the energy densities in powers of $u=\tau^{-2/3}$, \eqn{bjorken},  the different metric functions can be expanded in a power series in this variable. After introducing the new holographic coordinate $s=1/(r \tau^{1/3})$, a series solution of the Einstein equation can be found by expanding   
\begin{align}
\label{eq:uexpansion}
A(\tau,r) & = \sum_{i=0} u^{i} A_{i}(s)  , \\
b(\tau,r) & = \sum_{i=0} u^{i} b_{i}(s) ,\\
c(\tau,r) & = \sum_{i=0} u^{i} c_{i}(s) .
\end{align}
With this assumption, the solutions of Einstein's equations become a set of ordinary differential equations (ODE's) in $s$ that can be solved by imposing AdS asymptotics at $s\rightarrow 0$ and regularity at the horizon, which using reparametrisation invariance, can be set\footnote{
In the holographic calculation a value of $\Lambda$ is chosen by setting the zero of the metric function $A$ (the approximate apparent horizon) to occur at $s=1$. Similarly, $\tau$ in this section should be understood as the dimensionless combination $\Lambda \tau$.
}  at $s=1$.  
The equations of motions for Gauss-Bonnet gravity can also be solved by using the same expansion, which has been 
recently used to determined the first few (three) orders of the expansion in \cite{DiNunno:2017obv}. Imposing that AdS asymptotics and the the metric has a horizon at $s=1$ the leading order solution is given by 
\begin{align}
\label{eq:zerothOrder_1}
A_{0} & = \frac{1}{2 \lGB}\left( 1 - \sqrt{1-4 \lGB (1-s^4)} \right) \, ,\\ 
b_{0} & = -2\log\left(L_{c}\right)  \,,  \\
\label{eq:zerothOrder_2}
c_{0} &=-2\log\left(L_{c}\right)  \,,
\end{align}
which coincides with the black-brane metric in Gauss-Bonnet gravity expressed in the Eddington-Finkelstein gauge\footnote{
The expansion of the non-trivial prefactor of the $g_{yy}$ component in \eqn{eq:genBImetric} to leading order in $u$ must also be performed to obtain the black-brane metric.}. 
Recalling the definition of $s$, this leading order solution may be interpreted as a black-brane falling in the holographic direction at a rate given by the temporal change of the temperature scale, as defined in \eqn{eq:defTeff}. 
Starting from this solution, all higher orders can be computed by solving a set of subsequent linear ordinary differential equations; in Appendix \ref{GenSoln} we describe a formal solution to all orders which can be used to organise the perturbative solution. 

From this set of differential equations, the energy density can be computed by analysing the $s\rightarrow 0$ limit of the functions $A_i$. Given the boundary expansion \eqn{eq:boundaryexpansion} and using that $A_0(s=0)=1/L_c^2$, all metric coefficients $A_i$ for $i \ge 1$ must vanish at the origin. Comparing the expression of the holographic stress tensor, \eqn{eq:Tabgen} with the gradient expansion \eqn{eq:uexpansion}, the energy density of the dual field theory is given by 
\be
\label{eq:epsilon_series}
\epsilon=- \frac{N_c ^2}{2 \pi^2} \frac{(2L_{c}^{2}-1)}{L_{c}^{3}} \frac{3}{4} u^{2} \sum_{n=0} u^{n} \frac{1}{4!} \left. \frac{d^4}{ds^4} A_{n}(s) \right|_{s=0} \,. 
\ee
Comparing this with the expression for the equation of state \eqn{eq:eos}, we can determine the late proper time expansion of the effective temperature. Combining this expansion with the definition of the
 anisotropy function $R$, \eqn{eq:Rdef}, we can use the series to determine the coefficients $r_i$, as defined in \eqn{eq:Rseries}.

Following the procedure outlined above, 240 orders in the gradient expansion of the energy density  \eqn{bjorken}   for strongly coupled $\N=4$ SYM, were determined in \cite{Heller:2013fn}. 
In this work we have extended this computation up to 380 orders and extended it to fixed negative values of $\lGB$.
Since the Gauss-Bonnet equations of motion contain many more terms, the computation of higher order expansion coefficients becomes much more numerically demanding than for $\N=4$ SYM. Similarly, the presence of a singularity, increasingly close to the horizon as $\lGB$ becomes more negative also makes the numerical computation more challenging.
The analysis in this paper is based on the determination of $N_{\rm coefficients}=$ 94, 86, 80, 66 coefficients for $\lGB=-0.1,\,  -0.2, \, -0.5,\, -1$ respectively. 
These coefficients, as defined in \eqn{bjorken} and \eqn{eq:epsilon_series}  can be found in the arXiv submission on this paper.

\section{\label{sec:resurgence} Resurgence}
\begin{figure}[t]
\begin{centering}
  \includegraphics[width=0.8\linewidth]{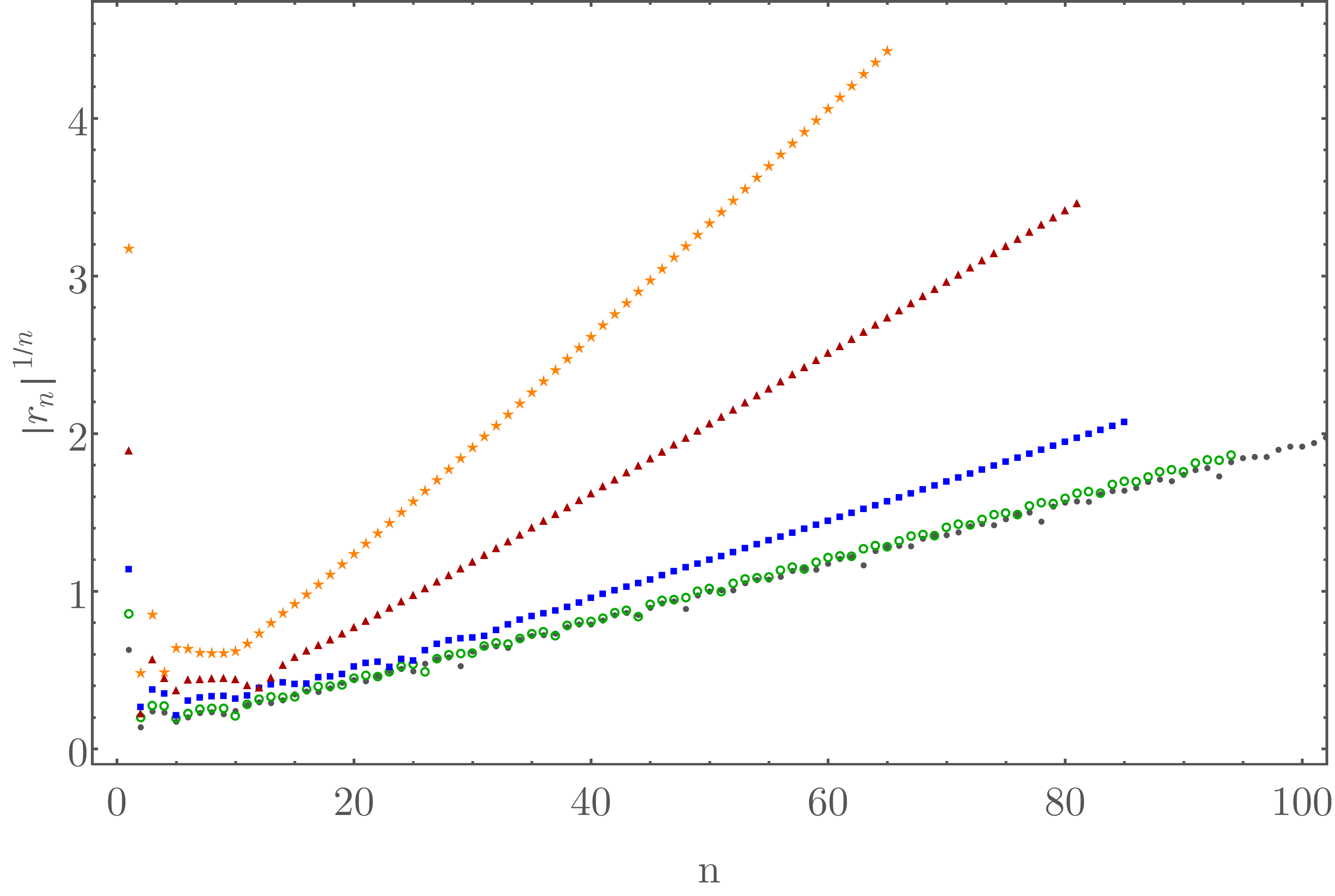}
  \caption{\label{fig_exp_growth} Behaviour of the series coefficients for the anisotropy function \eqn{eq:Rdef} as a function of the expansion order $n$ for $\lGB= 0$ (grey solid circles), $\lGB= -0.1$ (green open circles), $\lGB= -0.2$ (blue squares), $\lGB= -0.5$ (red triangles) and  $\lGB= -1$ (orange stars).
  }
  \end{centering}
\end{figure}

One of the main conclusions of the analysis of boost invariant flows is that the hydrodynamic expansion does not converge and behaves instead as an asymptotic series  \cite{Heller:2013fn}.
 This conclusion is based on the factorial growth of the coefficients in the large order hydrodynamic series. As a consequence, for any fixed gradient strength, increasing orders in the gradient expansion lead to larger contributions to the hydrodynamic functions. 
This behaviour has been observed both in $\N=4$ strongly coupled SYM   \cite{Heller:2013fn} and in kinetic theory in the RTA approximation  \cite{Heller:2016rtz}, as well as in phenomenological completions of hydrodynamics \cite{Heller:2015dha}. 
As expected, this same behaviour is also observed in Gauss-Bonnet gravity.  In \fig{fig_exp_growth} we show the growth of the magnitude of the series coefficients for the anisotropy function \eqn{eq:Rseries}.

The factorial growth of the series coefficients indicates that the hydrodynamic series may be Borel summable.  As is standard (see \cite{Dorigoni:2014hea,DuneLectures} for recent reviews in resurgence), we may define the series expansion of the Borel transform of the anisotropy function $R$, as
\be
\label{eq:BorelSeries}
R_{B} (\xi) = \sum_{n=1}^{N_{\rm coefficients}} \frac{r_n}{ n!} \xi^n \,. 
\ee
Unlike the hydrodynamic series, since the leading $n!$  growth has been removed, 
the Borel transform 
 defined above has a finite radius of convergence, which is controlled by the asymptotic large n slope of the growth of the coefficients shown in 
 \fig{fig_exp_growth}. The slopes of these curves grow as $\lGB$ becomes more negative, which in turns means that the radius of convergence of each series decreases with decreasing $\lGB$.  From the point of view of finite coupling corrections this may be  a natural expectation, since at finite coupling we expect the magnitude of the gradient corrections to grow. In the rest of this section we will explore the precise dynamics behind this observation.

The finite radius of convergence indicates that the Borel transform of the anisotropic function possesses singularities in the complex-$\xi$ plane.
The Borel transform and the original series are related via a Laplace transform. In order to be able to perform this integral, we will need to first analytically continue the Borel transform beyond its radius of convergence. A standard method to do this is to approximate the Borel transform via a Pad\'e approximant as
\be
R_B(\xi) \approx \mathcal{P}_{N,M}(\xi) =\frac{\sum^{N}_{j=0} n_i \xi^j}{1+\sum_{i=1}^M d_i \xi^i} \, ,
\ee
where all $N+M+1$ coefficients are fixed by demanding that the power series of the Pad\'e agrees with \eqn{eq:BorelSeries}. The choice of $N$ and $M$ are arbitrary, with the constraint that $N+M = N_{\rm coefficients}$. All the analysis of this work is based on symmetric Pad\'e approximants, with $N=M=N_{\rm coefficients}/2$, where we check the stability of our results by varying $N$.

\begin{figure}[t]

  \includegraphics[width=0.5\linewidth]{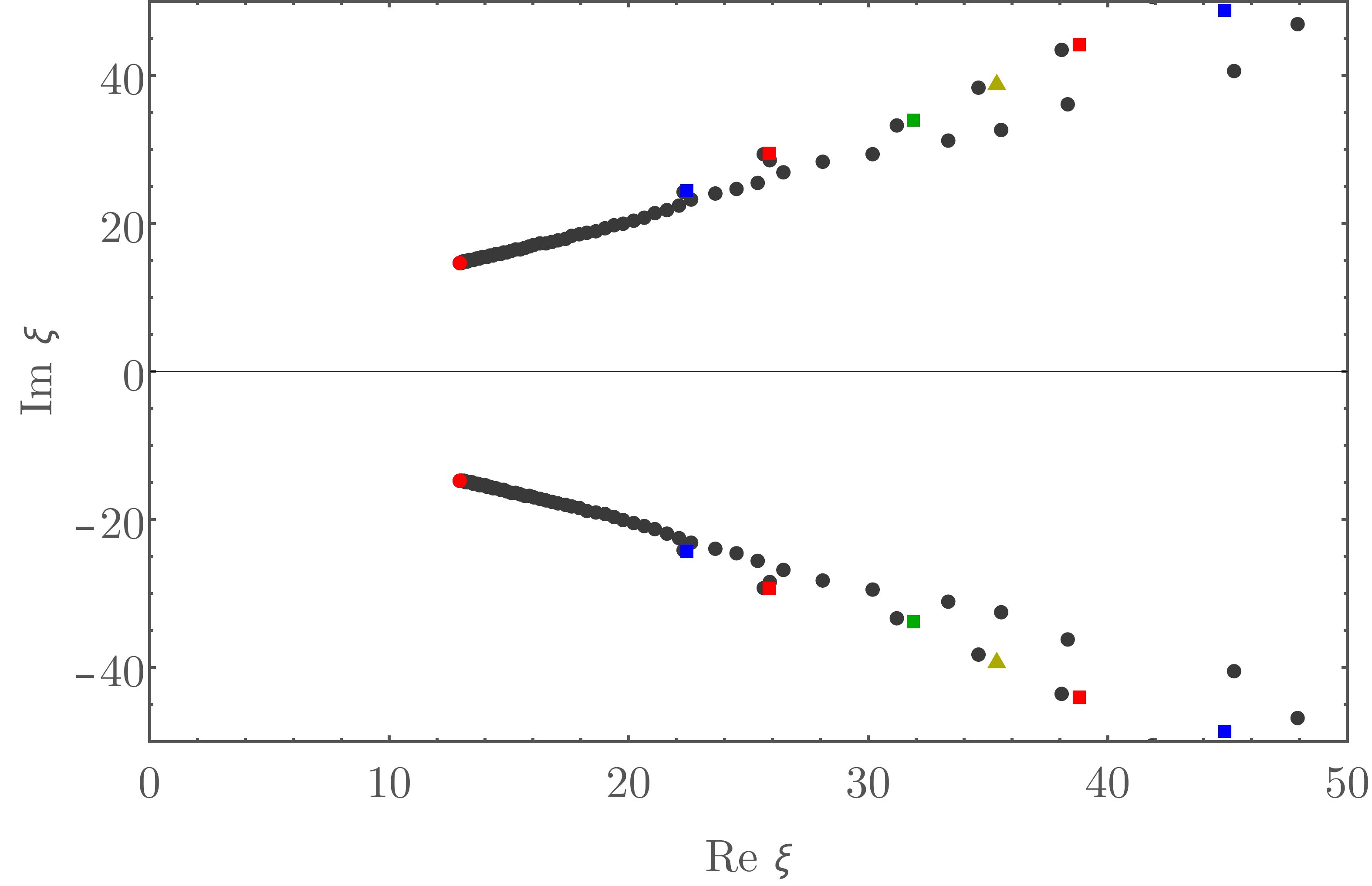}
  \put(-180,115){$\lambda_{GB}=0$}
  \includegraphics[width=0.5\linewidth]{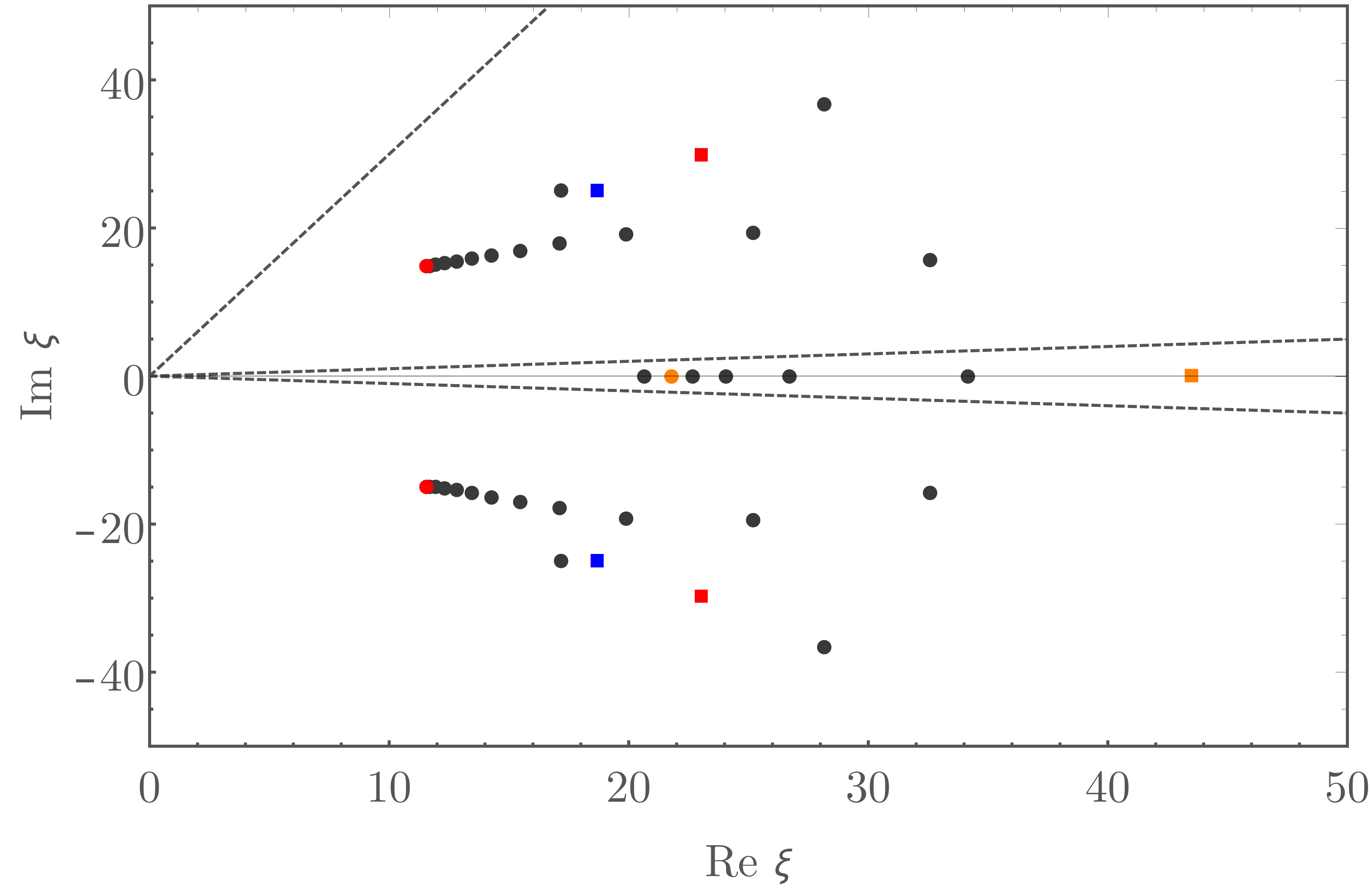}
  \put(-80,115){$\lambda_{GB}=-0.1$} \\
  \includegraphics[width=0.5\linewidth]{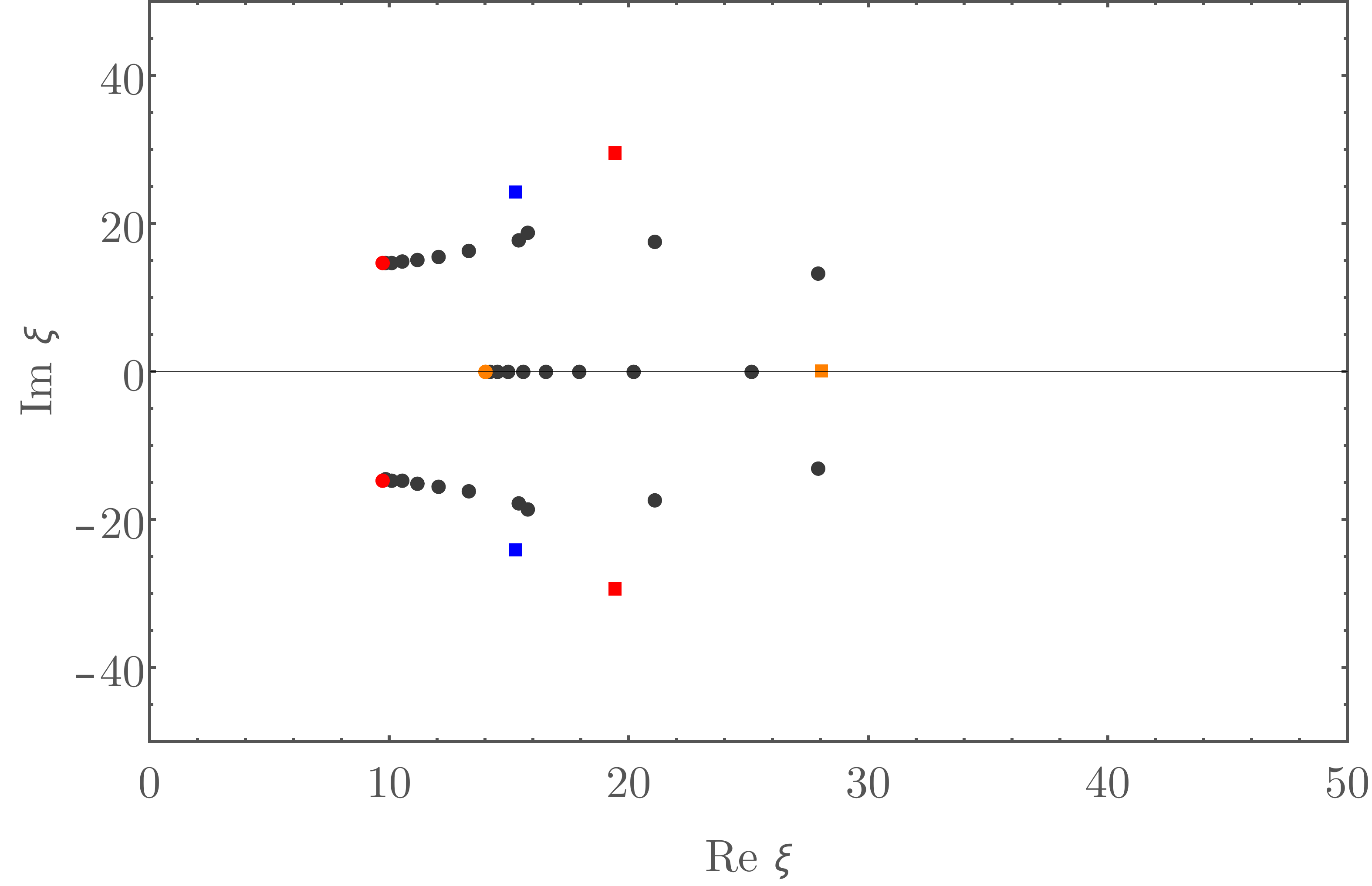}
    \put(-80,115){$\lambda_{GB}=-0.2$} 
  \includegraphics[width=0.5\linewidth]{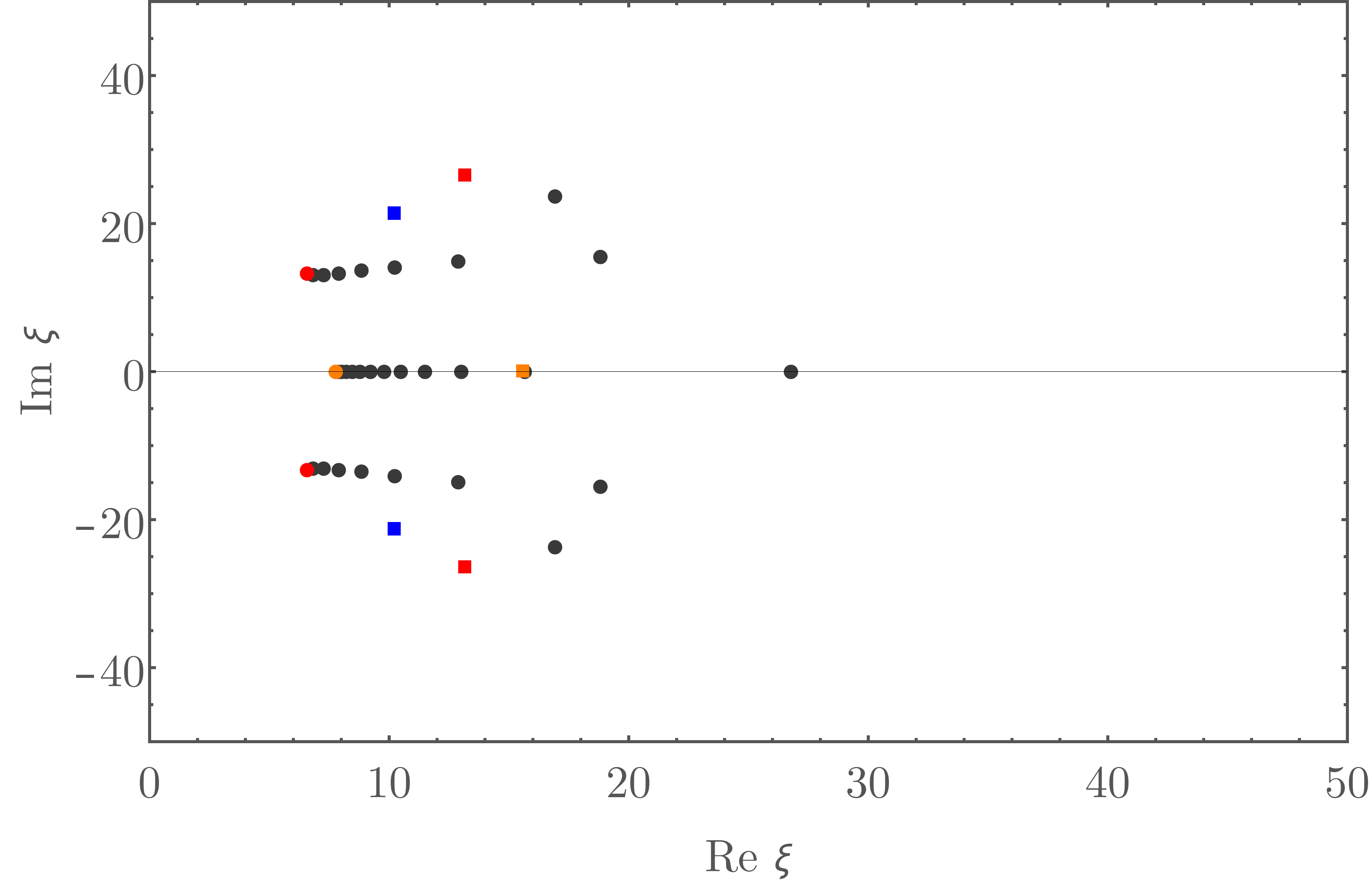}
    \put(-80,115){$\lambda_{GB}=-0.5$} \\
  \includegraphics[width=0.5\linewidth]{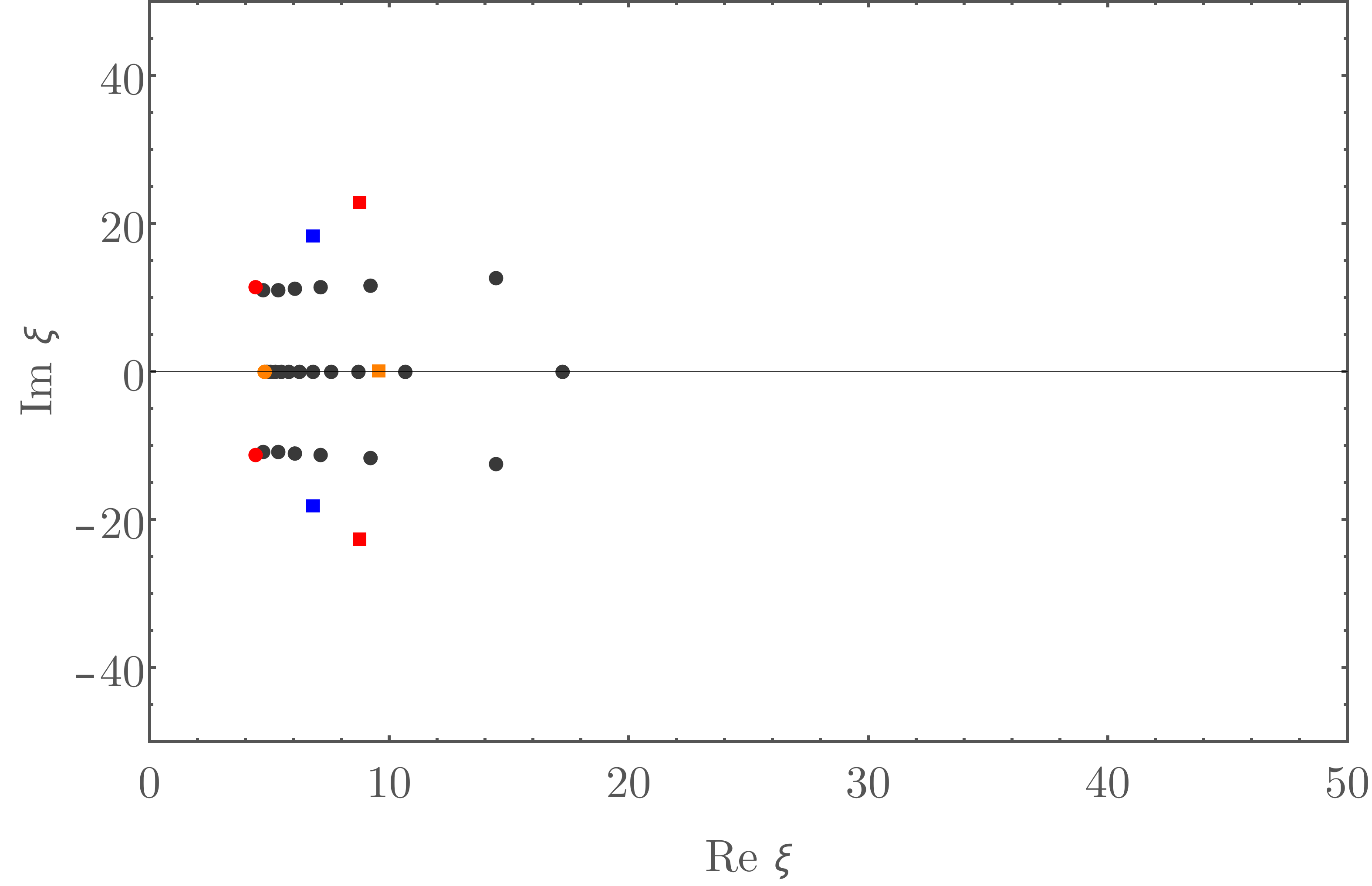}
      \put(-80,115){$\lambda_{GB}=-1$} 
 \includegraphics[width=0.5\linewidth]{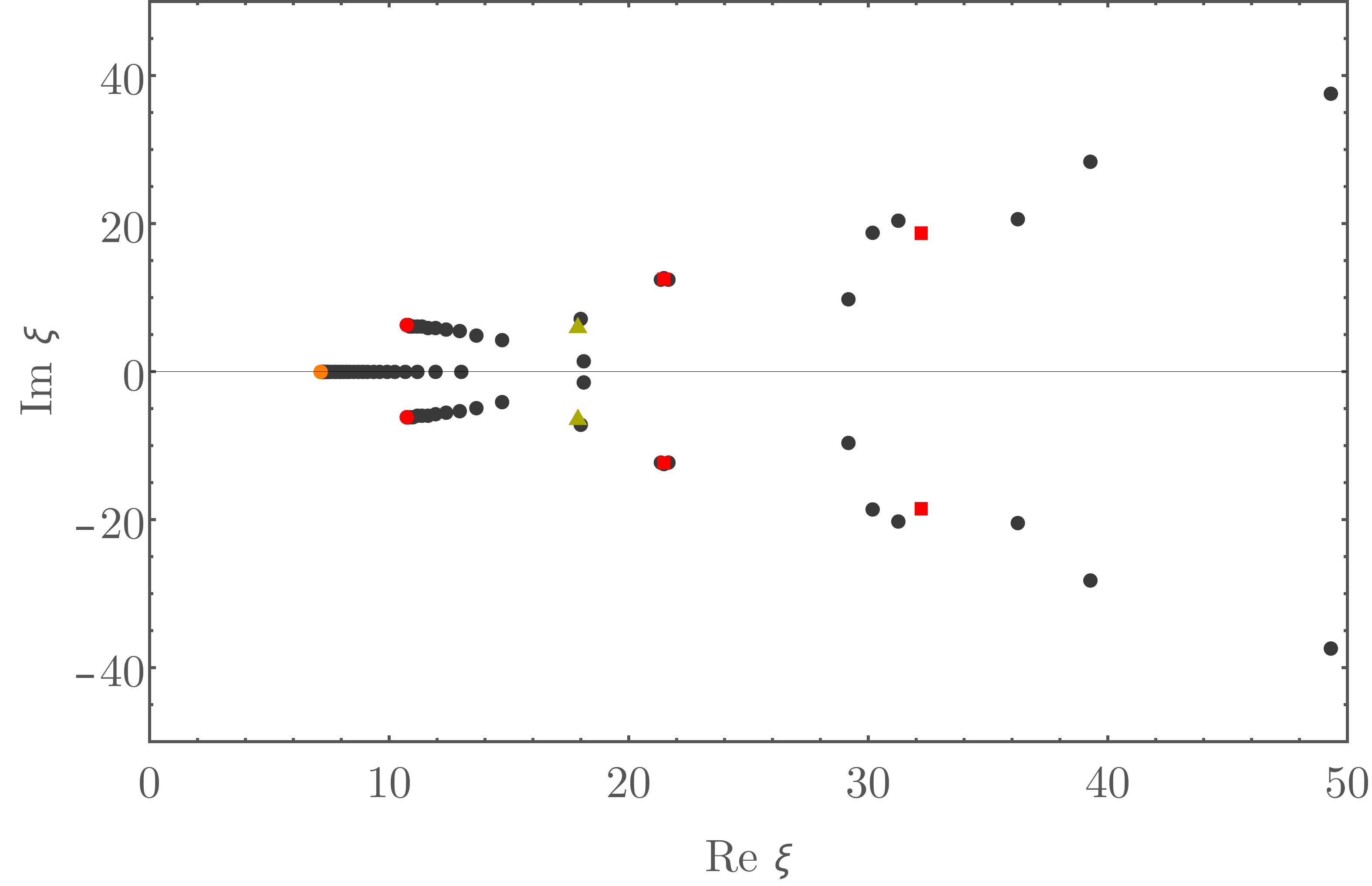}
       \put(-180,115){RTA Kinetic Theory} \\
  \caption{ \label{fig:BorelPlanes}
  The lower right panel is the Borel plane of kinetic theory in the RTA approximation using the coefficients computed in \cite{Heller:2016rtz}. For this plot we have chosen the product of the relaxation time times temperature, $\gamma\equiv \tau_R T=\pi/15$, so that shear viscosity of the RTA coincides with the $\lGB=-1$ value.
  The straight dashed lines in the upper left channel shows the contours of integration used in Section.~(\ref{sec:resum}), $\Cc_c$, $\Cc_+$ and $\Cc_-$ in decreasing slope order. Similar contours are used for all other cases. The solid grey circles indicate poles of the Pad\'e approximant of $R_{B}(\xi)$, the Borel transform associated with the anisotropy function $R(w)$ for different values of $\lGB$. The expected positions of singularities predicted from the quasi-normal mode frequencies closest to the origin, both for complex (red) and purely dissipative modes (orange), are marked by solid circles. All integer multiples of these frequencies are given by squares of the same colour. The subsequent $2^{\text{nd}}$ and $3^{\text{rd}}$ QNM frequencies are marked by blue and green squares. The modes that correspond to the sum of the $1^{\text{st}}$ and $2^{\text{nd}}$ QNM frequencies (in the $\mathcal{N}=4$ case) or the sum of the first two leading modes (in the RTA case) are given by a yellow triangle. 
  }
\end{figure}

One of the advantages of the Pad\'e approximant is the fact that, by construction, it allows for the emergence of poles in the complex $\xi$ plane, which can be related to the finite radius of convergence of the Borel transform. 
In \fig{fig:BorelPlanes} we show the positions of the poles of the Pad\'e approximant for different values of $\lGB$.
In the upper left panel we show the pole structure for $\N=4$ SYM as previously computed in \cite{Heller:2013fn,Florkowski:2017olj}, but including an additional 140 coefficients of the gradient series. The rest of the panels show our results for different negative values of $\lGB=-0.1,\, -0.2,\, -0.5, \, - 1$. For comparison,  in the lower right panel we show the Borel plane for the same analysis\footnote{Note that in \cite{Heller:2016rtz} the definition of the Borel transform was slightly different to ours, which implies that our results are not identical } in kinetic theory within the RTA, using the coefficients tabulated in \cite{Heller:2016rtz}. For all these cases additional poles exist for very large values of $\left| \xi \right | $; however these have a strong dependence on the Pad\'e order, which indicates that they are numerical artefacts.

The singularity structure of the Borel Plane is particularly interesting. As a first observation, the locations of the poles control the convergence of the Borel series, since the distance of the closest pole to the origin is proportional to the inverse of the slope of coefficients shown in \fig{fig_exp_growth}.
Furthermore, in all cases, the Pad\'e approximant exibits an accumulation of alternating poles and zeroes, starting at a well defined point in the borel plane. This concentrated sum of simple poles indicates the emergence of a branch cut
\cite{2013arXiv1308.4453Y}.
Nevertheless, the structure of poles at finite $\lGB$ is qualitatively different to that of $\N=4$ SYM at infinite coupling. While in the latter case all poles are complex, for all finite $\lambda_{GB}$ new branch cuts emerge along the real axis. For small negative $\lambda_{GB}$ these new branch cuts are far from the origin, but as $\lambda_{GB}$ becomes more negative these poles move closer to $\xi=0$, and eventually dominate the radius of convergence for the Borel Transform. This behaviour qualitatively interpolates the structure of the infinitely coupled limit of $\N=4$ SYM with the expectation from perturbation theory as obtained via kinetic theory in the RTA approximation.

We can note from Fig. (\ref{fig_exp_growth}) that for $\lambda_{GB} = 0$ and $\lambda_{GB} = -0.1$ the leading behaviour of the coefficients at large $n$ follows the form of an oscillating factorial function ($r_n\sim n! \cos(a n))$, in a similar fashion to scenarios noted in \cite{Spalinski:2017mel,Aniceto:2015mto}. As we vary $\lambda_{GB}$ to decreasing values we find that the oscillating behaviour becomes suppressed and the coefficients tend to follow $r_n\sim n!$ as in \cite{Heller:2016rtz}. This is consistent with the dominant contribution to the large $n$ coefficients for the hydrodynamic series transitioning from two dispersive non-hydrodynamic modes, to a single dissipative non-hydrodynamic mode.

    Having analytically continued the Borel transform beyond the power expansion, we can determine the anisotropy function beyond the power series via the inverse Borel transform
\be
\label{eq:Rborelback}
R(w)=  w \int_{\Cc} d\xi \,  e^{-w \xi} R_B(\xi) \,,
\ee
where $\Cc$ is a contour in complex plane which connects $\xi=0$ and $\xi=\infty$. The presence of singularities in $R_B$ shows that different choices of contour $\Cc$ yield different answers. 
Since  we require this to be an analytic continuation for every complex value of $\omega$ or $\xi$, this implies that all choices of $\Cc$ must yield identical results. The theory of resurgence indicates that the anisotropy function cannot be simply approximated by a gradient expansion, but must also incorporate non-perturbative contributions in the gradient strength. Denoting by $\xi_0^{(\alpha)}$ the origin of each independent branch cut in $R_B$ (each of which lead to an independent trans monomial) this trans-series is given by \cite{Heller:2013fn,Aniceto:2015mto,Florkowski:2017olj},
\be
\label{eq:trans}
R(\omega)=\prod_{\alpha=1}^{N}\left(\sum_{n_{\alpha}=0}^{\infty}\Omega_{\alpha}^{n_{\alpha}}\right)\Phi_{(n_{1}...,n_{N})}(\omega),
\ee
where $N$ denotes the number of independent non-perturbative modes, and 
 the functions $  \Phi_{(n_1 \, n_2\, ... ,n_{N})} (w)$  
 admit power series  in inverse powers of $w$ at large $w$. The non-perturbative behaviour in the gradient strength is encoded in the functions $\Omega_{\alpha}$, given by 
\be
\label{eq:Omegadef}
\Omega_{\alpha} =  C_{\alpha}w^{\gamma^{(\alpha)}} e^{- \xi_0^{(\alpha)} w} \,, 
\ee
where $\gamma^{(\alpha)}$ is constant for each branch cut which may be determined from the analysis of residues along the branch cut and $C_{\alpha}$ are 
Stokes parameters, which must be chosen such that the non-perturbative  ambiguity obtained in the Borel-summation of $\Phi_0 (w)$, is exactly cancelled by the next terms in the trans-series, yielding  a real final result \cite{Dorigoni:2014hea,Aniceto:2015mto}. 
These parameters will jump discontinuously every time the contour $\cal C$ crosses a singularity in the Borel plane. However, this reality condition does not completely fix these complex numbers \cite{Aniceto:2015mto}.
While in the hydrodynamic limit all the information about initial conditions reduces to an overall scale, the different values of these constants correspond to additional information on the initial state of the evolution, which controls the magnitude of the non-perturbative modes. 

As noted in 
\cite{Heller:2013fn}
the form of this trans-series coincides with the expected contribution of the evolution from non-hydrodynamic perturbations of the system away from local equilibrium in a boost invariant expanding medium of the equilibrium state.
As it is well known, at strong coupling these non-hydrodynamic excitations are characterised by a set of characteristic complex frequencies, which in the dual theory coincided with the quasi-normal modes of the associated black-branes. In the adiabatic approximation, each of these these excitations relax according to the local relaxation
\be
\delta R^{(\alpha)} \sim {\rm exp} \left\{i \int \omega^{(\alpha)} (\tau) d\tau\right\} \,,
\ee
where $\omega^{(\alpha)}$ is the characteristic frequency of each mode. Since the system under consideration is conformal, the $\tau$-dependence of those frequencies is controlled by the effective  local temperature.
From the late time T-dependence, the position of the branch cut can be related 
the frequencies of the quasi-normal modes as 
$\xi^{(\alpha)}_0=2 \, i \,  \omega^{(\alpha)}_{\rm QNM}/3$.
 The emergence of these characteristic frequencies in the expanding case can be found explicitly by searching for power series solutions of the form \eqn{eq:uexpansion} supplemented by non-perturbative pre-factors 
\begin{align}
\label{eq:uexpansionNP}
\delta A^{(\alpha)}(\tau,r) & = \Omega_{\alpha}  \sum_{i=0} u^{i} \delta A^{(\alpha)}_{i}(s)  , \\
\delta b^{(\alpha)}(\tau,r) & =\Omega_{\alpha}  \sum_{i=0} u^{i} \delta b^{(\alpha)}_{i}(s) ,\\
\delta c^{(\alpha)}(\tau,r) & = \Omega_{\alpha}  \sum_{i=0} u^{i} \delta c^{(\alpha)}_{i}(s).
\end{align}
We have checked that even at finite $\lGB$, at leading order in gradients, the resulting ODEs become independent of $\gamma^{(\alpha)}$ and coincide with the QNM equations of the static black-brane in \cite{Grozdanov:2016vgg}, after the appropriate relation of $\xi^{(\alpha)}_0$ with the quasi-normal mode frequency\footnote{We thank M. Spali{\'n}ski for useful discussion on this point.}.

From the above result, the observed qualitative differences between the Borel planes of $\N=4$ SYM and Gauss-Bonnet gravity can be traced back to the structure of quasi-normal modes. As noted in \cite{Grozdanov:2016vgg}, this higher-derivative theory possesses a new set of dissipative (imaginary) quasi-normal modes
in addition to the characteristic discrete complex modes of $\N=4$ SYM. These purely imaginary poles are not an artifact of this particular higher-derivative theory. 
As explicitly shown in  \cite{Grozdanov:2016vgg}, the higher-derivative term responsible for finite coupling corrections to $\N=4$ SYM also lead to this new type of relaxation mode; and the t'Hooft coupling dependence of these poles is qualitatively similar to the $\lGB$ dependence as long as $\lGB$ is negative. We can therefore infer that the structure of the Borel plane singularities for $\N=4$ will also be qualititatively similar to the one observed in our analysis. It is rewarding to realise that these corrections seem to interpolate between the perturbatively weak and infinitely strong coupling limits. 

To explicitly show the relation between the quasi-normal mode spectrum and the Borel plane singularities we show the positions of these characteristic frequencies, after an appropriate rescaling, in \fig{fig:BorelPlanes}. In this figure the positions of the singularities associated with the first purely imaginary and complex QNM's (with smallest imaginary part, i. e. the smallest damping rate) are shown by the orange and red solid circles respectively. Note that from the relation above between the QNM frequency and the parameter $\xi_{0}^{(\alpha)}$ that these correspond to poles in the Borel plane with the smallest real part. In all panels, such poles coincide with the start of an accumulation of singularities in the Pad\'e approximant, which may be interpreted as the origin of the branch cut. 
The singularities associated with higher QNM's must also be present in the Borel plane; in \fig{fig:BorelPlanes} we have shown the positions of these singularities associated with the second and third complex QNM's (with the next two smallest imaginary parts) by blue and green squares. Integer mutliples of all QNM frequencies above are given in squares of the same associated colour.

In all our finite coupling computations we do not identify poles coinciding with higher order modes. This however is likely an artefact of the limited number of coefficients we have been able to extract from our computations. For $\N=4$, where we are able to determine many more coefficients, these singularities indeed emerge, as already pointed out in \cite{Florkowski:2017olj}. Note  that resonant singularities, associated to the product of trans-monomials in \eqn{eq:trans} are also visible; in Fig. (\ref{fig:BorelPlanes}), yellow triangles are used to identify the sum of the lowest two complex QNM frequencies (in the case of $\N=4$). We conclude this description by noting that these resonant singularities are also visible in the Borel plane of the RTA kinetic theory; this observation strengthens the significance of these structures, which have only been observed in the non-linear analysis of \cite{Heller:2016rtz}, but do not correspond to poles of the retarded correlator of stress tensor in the linear response analysis of RTA kinetic theory performed in  \cite{Romatschke:2015gic}. 

\section{\label{sec:resum} Resummations  and the Hydrodynamic Attractor}

We now study the extension of the anisotropy function for small values of $w$ by analysing the result of the inverse Borel transform, \eqn{eq:Rborelback}. As we stressed in the previous section the presence of poles in the Pad\'e approximant, $R_B(\xi)$, will lead to an ambiguity of the inverse Borel transform, since depending on the choice of contour we will obtain different answers. This ambiguity can be fixed by demanding that the coefficients $C_{\alpha}$ (Stokes parameters) will be discontinuous across each branch cut, as is known as Stokes phenomenon. However, as already mentioned, this procedure does not  completely fix the value of these coefficients on a given contour, only its change across Stokes lines. The remaining  ambiguity can only be determined from additional knowledge of the far-from-equilibrium early time dynamics of the system's evolution, since specifying different values is equivalent to selecting different choices of initial configurations for the evolution. 

Among all the different possible time evolutions of the system, it has been recently proposed that there is a particular configurations which behave as an attractor in the space of initial conditions \cite{Heller:2015dha}. While currently there is not a precise definition of the attractor (see \cite{Behtash:2017wqg} for a recent attempt to provide such a definition based on the theory of non-autonomous dynamical systems), numerical analysis of different theories have identified well defined 
attractor solutions at all $w$, at which all time evolutions of the system converge. 
Given the previously mentioned difficulties in studying the very early time dynamics of this high-derivative gravity, in this section we will constraint the properties of the attractor solution in holographic Gauss-Bonnet by resumming the hydrodynamic series.

An obvious physical requirement for the choice of contour is that  $R$ must be real. In the case of $\N=4$ SYM this requirement is easily fulfilled by choosing, for example,  the real axis as an integration contour, while setting all the coefficients $C_{\alpha}$ to zero. This choice was recently analysed in 
\cite{Spalinski:2017mel}, motivated by the results obtained in a hydrodynamic theory with a similar singularity structure in the Borel plane as that of $\N=4$ \cite{Aniceto:2015mto}. In that theory, the direct integration of the inverse Borel transform in the real axis coincided with the numerically computed attractor for  of $w>0.3$.
We have tested that the additional coefficients computed in this work do not change the result of this resummation in this case. 
Recent analysis of exact solutions of IS hydrodynamics has shown \cite{Denicol:2017lxn} that in that model the attractor coincides with the direct resummation of the hydrodynamic series explicitly setting to zero all non-perturbative tails. 

 For finite $\lGB$,  the presence of poles in the real axis complicates the extraction of the anisotropy function, since any contour that avoids those poles generically 
 leads to an complex $R$. This feature is a clear manifestation of the need to include non-perturbative corrections in the form of a trans-series to determine the anisotropy function beyond the power expansion. trans-series corrections were indeed studied in \cite{Heller:2015dha} in the context of BRSSS hydrodynamics, which also exhibit real poles in the Borel plane. While the expansion \eqn{eq:uexpansionNP} provides a clear starting point to complete this program in holographic theories, computing these corrections are numerically more challenging and goes beyond the scope of this work. Nevertheless, since these corrections are exponentially suppressed at large $w$ we will use the leading term in the trans-series to constraint the dynamics of the resummation.
 
In this work we focus on the real part of the anisotropy function computed by performing the inverse Borel transform, \eqn{eq:Rborelback} of the Pad\'e approximant determined in the previous section. We choose a contour of integration given by a straight line in the complex plane $z=\xi e^{\theta}$, with $\theta>0$ such that all the poles with positive imaginary part lie above the contour.  For latter reference, we will refer to this contour as $\Cc_+$ and it is shown in the upper right panel of \fig{fig:BorelPlanes}. As already mentioned this choice leads to a complex $R$-value; however, as expected, at large $w$ this imaginary part becomes increasingly small. Note that by choosing the real part we are making the computed $R$ value independent of whether the integration is performed 
 along $\Cc_+$ or along an analogous contour $\Cc_-$ obtained by reflexion along the real axis (see \fig{fig:BorelPlanes}), 
since the discontinuity across the real axis is purely imaginary. In essence, by this prescription we are effectively incorporating part of the first trans-series corrections. In fact, a procedure to determine these coefficients is precisely to choose the $C_{\alpha}$ to cancel all imaginary contributions. 
However, this is not the full answer, since the functions 
$\Phi_{(n_{1}...,n_{N})}$ 
in \eqn{eq:trans} may contain additional imaginary parts which are cancelled only by higher order terms in the trans-series either associated to independent real poles further away from the origin (absent in Gauss-Bonnet) or by integer multiples of the leading real pole. Nevertheless, these contributions possess stronger exponential suppression factors, which make them only relevant at sufficiently small time. Note that this prescription implies, in particular, that all the trans-series coefficients $\Cc_\alpha$ associated to complex singularies are set to zero along this integration contour. 
 
 \begin{figure}[t]
  \includegraphics[width=0.5\linewidth]{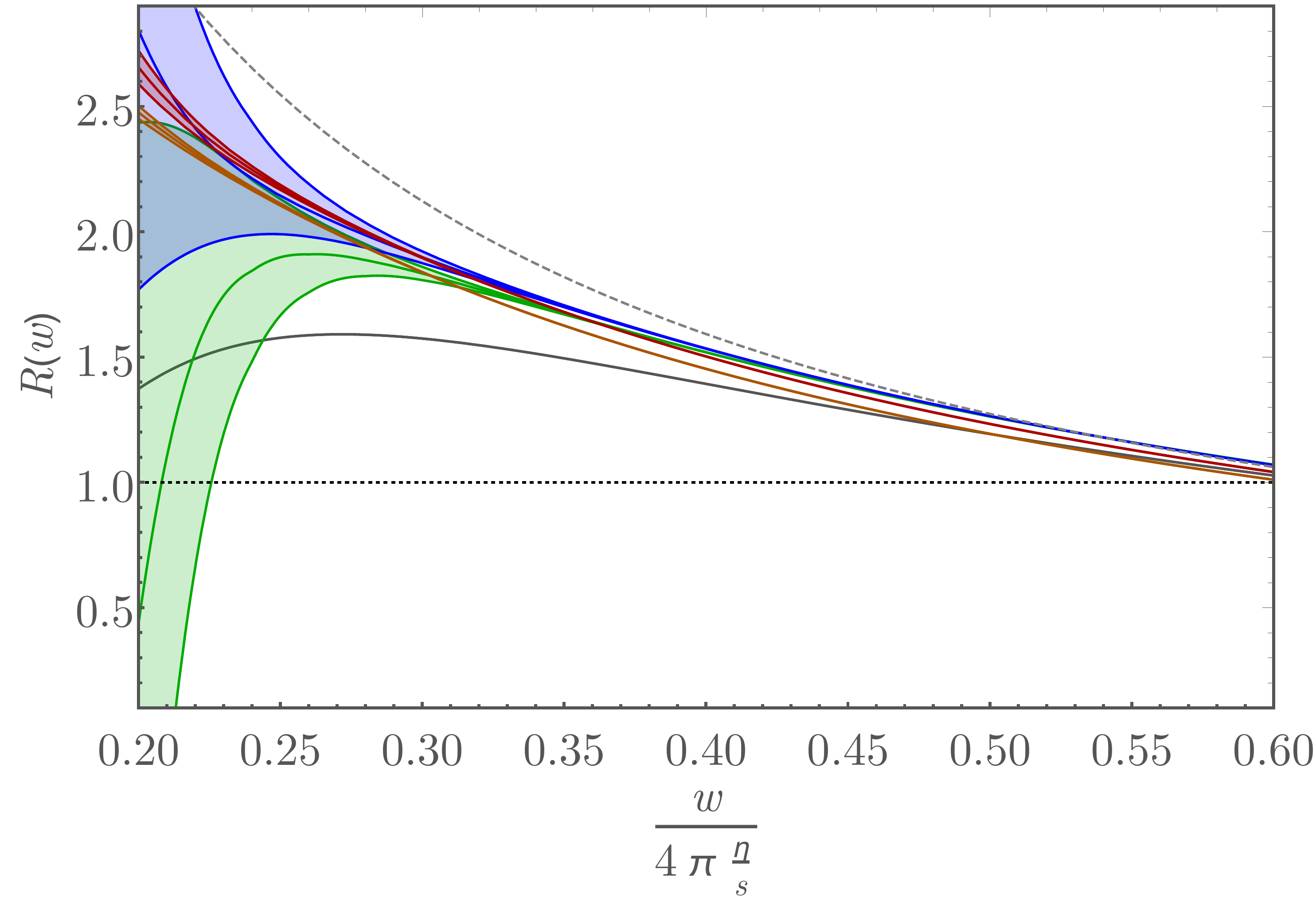}
    \includegraphics[width=0.5\linewidth]{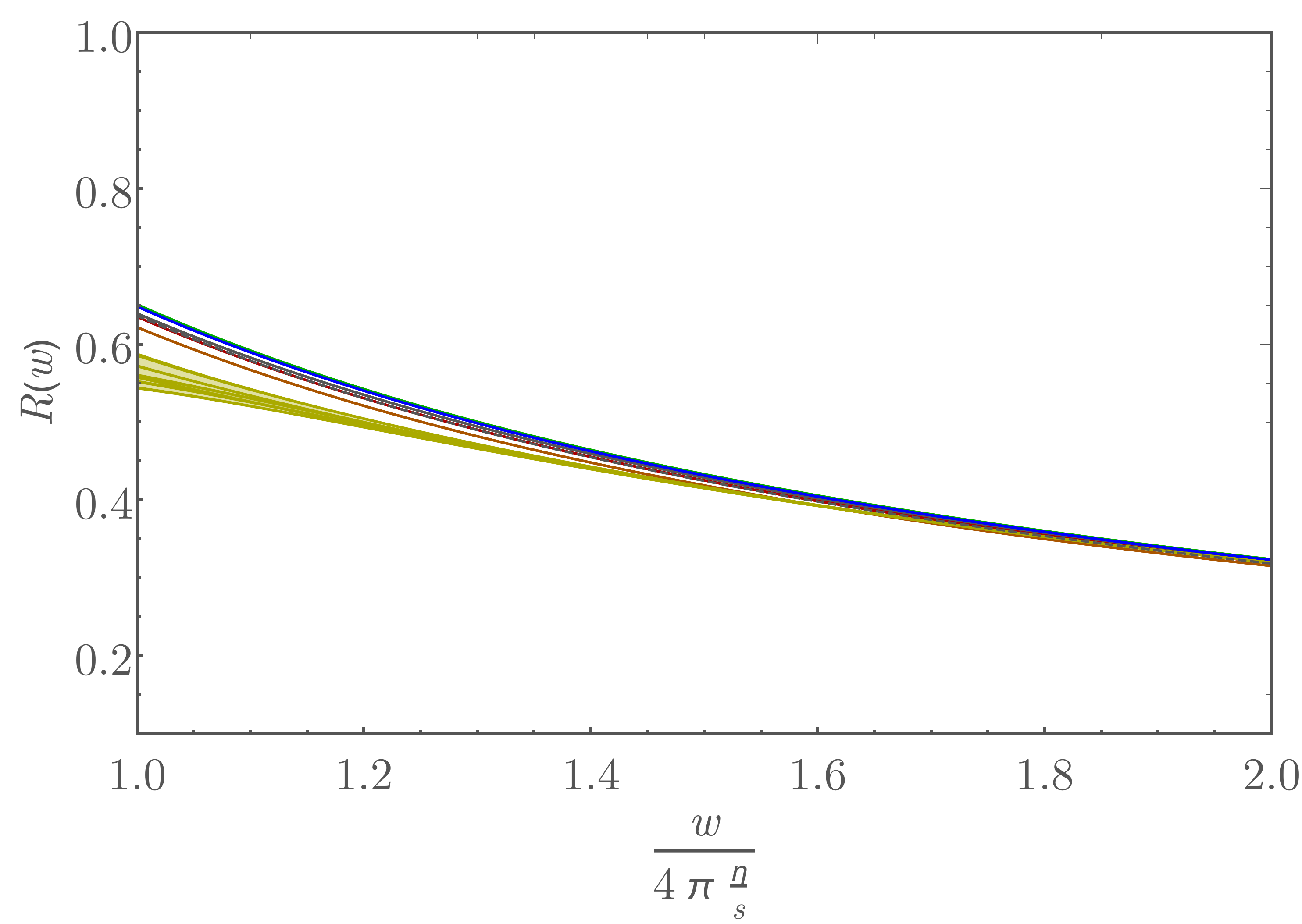}
      \caption{\label{fig:Rresummed} Anisotropy function for different values of $\lGB$ (left) and RTA (right) as a function of rescaled gradient $w s / 4\pi \eta$. 
      The  grey, green, blue, red and orange  curves correspond to the real parts of the inverse Borel transform of the leading order in the trans-series while the yellow curve in the left panel corresponds to RTA (for different choices of the Pad\'e order $N$). The bands are determined by adding and subtracting the imaginary part of the inverse Borel transform, as a gauge of the importance of additional trans-series contributions. In both panels, the dashed line corresponds to the first order hydrodynamic prediction $R^{\rm 1st}_{\rm hyd}$.
The grey, red and orange curves show no notizable deviation for the entire range plotted above. The green and blue curves are stable with respect to the choice of $N$ for $\frac{w}{4\pi \frac{\eta}{s}}> 0.25$, with deviations remaining within the same order of magnitude as the curves plotted above. The sensitivity of the RTA bands to different choices of $N$ is no greater than $6 \%$ for  $\frac{w}{4\pi \frac{\eta}{s}}>1$.
}
\end{figure}

The results of this integration for all the different values of $\lGB=0,\, -0.1.\, -0.2,\, -0.5, \, -1$ are given by the grey, green, blue, red and orange curves displayed in the left panel of \fig{fig:Rresummed}. 
For all non-zero values of $\lGB$  we have supplemented each curve with a band generated by adding and subtracting to the real part the imaginary part of the integral \eqn{eq:Rborelback}.  
When the band is narrow, this is a conservative estimate of the deviation of the trans-series from our prescription, since, as already argued, additional contributions are exponentially suppressed. As the width of the band increases, the sensitivity to the Pad\'e order and the number of coefficients also increases. 
 In the right panel of  \fig{fig:Rresummed} we compare the results from our holographic computation to the resummation of the RTA kinetic theory coefficients
 from \cite{Heller:2016rtz}.
 Following \cite{Keegan:2015avk}, to better compare the different theories we have rescaled the values of $w$ by the viscosity to entropy density ratio, such that the first order hydrodynamic prediction  $R^{\rm 1st}_{\rm hyd}$, shown by the dashed line, agrees in all theories by construction. Even though the RTA computation is performed with 200 coefficients of the hydrodynamics series, we find that the inverse Borel transform is much more sensitive to the Pad\'e order, which prevents us from studying the very small $w s/\eta$ regime.
 
 The inspection of this figure shows that, after $w$ is properly rescaled, the evolution of the anisotropy function is very similar, but not identical, for all cases considered, at least at sufficiently large values of $w$. All the resummations at fixed $\lGB$ exhibit small imaginary parts for $w\lsim 4\pi \eta/s$, a region where viscous corrections are large. Note also that the magnitude of the non-perturbaive corrections does not scale with the ratio of $\eta/s$, since the width of the different bands at similar values of the re-scaled variable is different. From the point of view of the gravitational dual, this is a consequence of the fact that the imaginary part of the non-hydrodynamic QNM's does not scale with the transport coefficients, at least for the values of $\lGB$ considered. Similarly, the effect of non-perturbative modes is bigger for the RTA calculation, and significant deviations from the leading order term in the trans-series persist at $w\gsim 4\pi \eta/s$.

When the width of the bands is small, we can use the resummation to explore the process of hydrodynamization of the system. As already observed in \cite{Spalinski:2017mel} for $\N=4$, the result of these resummations quickly approaches the first order hydrodynamic predictions for all the values of $\lGB$. 
 To better quantify this process, we will assume that the system has hydrodynamized at $w_{\rm hyd}$ if for any larger value of $w$ the anisotropy function satisfies
\be
\frac{\left|R-R_{\rm hyd}\right|}{R_{\rm hyd}} <0.1\,,
\ee
 where $R_{\rm hyd}$ is the first order hydrodynamic expression \eqn{eq:Rseries}. 
 The values of $w_{\rm hyd}$ and the corresponding anisotropy for the different theories are tabulated in Table~(\ref{table}). 
 As $\lGB$ becomes more negative, the value of the temperature-normalised gradient $w$ at which hydrodyninamization occurs increases, as expected by the fact that the dual fluid is more viscous. Nevertheless, 
 as in other theories where the resummation has been performed \cite{Heller:2015dha,Spalinski:2017mel}, $R_{\rm hyd}$ approximates the resummed result even when the value of this normalised gradient is comparable to the microscopic scale.
  At these small values of the inverse gradient, the anisotropy function is larger than 1, which means that the viscous contribution to the pressures is as large as the equilibrium pressure, demonstrating that the contribution of higher order terms is potentially large. 
   This is once again a manifestation of hydrodynamization without isotropization as discussed in \cite{Heller:2011ju}. 
  In fact, since the series is only asymptotic, it is easy to test that the corrections given by the truncated hydrodynamic series at orders greater than 10 give divergent and sign alternating contributions at those values of $w$. These corrections are, nevertheless, tamed by the resummation.

In this table we have also quoted the values obtained for RTA. As already mentioned, these results are much more sensitive to the Pad\'e order and the extracted values reflect this sensitivity. This sensitivity hints towards a larger contribution of the non-perturbative corrections, which we will explore below in detail, making the hydrodynamization interpretation harder. Nevertheless, it is worth noting that all the computations performed via the gauge/gravity duality hydrodynamize at comparable values of the viscosity re-scaled gradient $w s/ \eta$, and significantly earlier than in RTA kinetic theory.
 Since both RTA and $\lGB$ may be viewed as oversimplified treatments of finite coupling effects in gauge theories, it would be interesting to investigate more realistic higher derivative corrections and collision kernels to explore whether the size of the re-scaled gradient at hydrodynamization shows consistent trends in these complementary approaches towards gauge theories at intermediate coupling.

 \begin{table}[t]
\begin{center}
\begin{tabular}{|c|c|c|c|c|c||c|}
\hline
$\lGB$ & 0 & -0.1 & -0.2 & -0.5 & -1 & RTA \\
\hline
$w_{\rm hyd}$ &0.43 & 0.46 & 0.56 & 0.93 & 1.85& 2.5 - 2.8
\\
\hline
$\frac {w_{\rm hyd}s }{4 \pi \eta}$  & 0.43 & 0.33 & 0.31 & 0.31 & 0.37 & 1.0 - 1.1
\\
\hline
$\left . R \right|_{w_{\rm hyd}} $& 1.33 & 1.74 & 1.87 & 1.85 & 1.57 & 0.55 - 0.57 \\
\hline
\end{tabular}
 \end{center}
 \caption{\label{table} Inverse gradient size and anisotropy function at hydrodynamization for different theories. Note that for RTA the quoted range reflects the sensitivity of the resummation to Pad\'e order and does not include the uncertainty associated with the imaginary part of the inverse Borel transform. 
  }
\end{table}

We now turn to the relation between the resummation of the hydrodynamic series and the attractor. As we have stressed, in performing our resummations we have implicitly selected some particular values of the initial conditions, which tantamount to a specific selection of the constants $C_{\alpha}$ in \eqn{eq:trans}. It is therefore unclear whether this choice leads to the hydrodynamic attractor.
In the simpler example of \cite{Heller:2015dha}, where the trans-series program has been performed, non-trivial values for this constant, beyond the cancellation of imaginary parts, must be introduced (fitted) to describe the numerically computed attractor.
 Therefore, to fully determine the attractor numerical computations from an early initial time are needed. 
 For $\N=4$, an attractor has been identified by studying the behaviour of different initial conditions at a very early initial proper time in \cite{Romatschke:2017vte}. As already stressed, performing these types of computations in Gauss-Bonnet holography introduced practical and conceptual difficulties which make them a challenge beyond the scope of this paper. 
 For this reason, in this paper we will use information extracted from the resummation to constrain the position of the attractor. 

To estimate the relaxation of different sets of initial conditions, we will focus on the dynamics of the leading non-perturbative corrections. From the point of view of holography, these may be interpreted as the effect of the least damped quasi-normal modes. As we have already mentioned, the $w$-dependence of these contributions could be obtained via the computation of a series expansion in gradients, analogous to \eqn{eq:uexpansion}, but supplemented with the non-perturbative prefactor $\Omega_\alpha$ for each mode.
 However, we can also determine the late time behaviour of this contribution by examining the discontinuities of the inverse Borel transform for different choices of the contour integration. 
  By inspection of the Borel planes at finite $\lGB$, \fig{fig:BorelPlanes}, we identify three representative contours of integration, which yield different answers for the inverse Borel transform. We have already used one of those contours,  $\Cc_{+}$, to define the inverse Borel transform above.  The second contour $\Cc_{-}$, is the reflection of  the previous to the lower half plane. Finally, the third contour is a straight line in the upper half plane at angle above the argument of the start of the complex branch cut, $\mathcal{C}_{c}$. All these contours are shown in \fig{fig:BorelPlanes}. Denoting by $R_{+}$, $R_{-}$ and $R_c$ the results of integrating \eqn{eq:Rborelback} over each of these contours, we define the discontinuities
 \be
 \label{eq:disc_def}
 i D_\pm (w) = R_{+} -R_{-} , \quad D_{c} (w) = R_{c}-R_{+}.
 \ee
$D_\pm$ is real and coincides with the imaginary part of $R_+$ while $D_{c}$ is complex. Note that we could have defined an equivalent discontinuity by reflecting both $\Cc_c$ and $\Cc_{+}$ to the lower half; however, this discontinuity is simply the complex conjugate of $D_c(w)$. 

Independence of the inverse Borel transform on the integration contour imposes that the constants $C_{\alpha}$ must differ for each integration. 
 Therefore, if the trans-series would contain only the contribution of the independent singularities $\xi_\alpha$ with smallest damping rate,
 this contribution would be proportional to the discontinuity, so that the ambiguity could be cancelled. 
 In more general cases the presence of additional non-perturbative modes as well as resonant contributions among poles imply that the cancellation is more subtle and the whole trans-series is necessary.
For this reason, the discontinuities we have computed are not solely dependent on the leading singularity. But since those additional contributions occur at larger values of the conjugate variable $\xi$, the exponential suppression of these contributions at sufficiently large values of $w$, $w>1/\xi$, is larger than those of the leading singularity, as inferred from the exponential contribution $\Omega_\alpha$ in \eqn{eq:trans}.

Within the above approximation, the late time behaviour of the different initial conditions is given by 
\be
\label{eq:ICR}
R_{IC} = R + a_1 D_{\pm} + a_2 {\rm Re} \left[D_c\right] +a_3 {\rm Im} \left[D_c\right] \,,
\ee
where R is the result of the resummation described above and the coefficients $a_i$ contain information from the early proper time evolution of the system beyond the hydrodynamic approximation. The functional form of each of these discontinuities is shown in Appendix~\ref{sec:npm}. Note that the edges of the band displayed in \fig{fig:Rresummed} corresponds to setting $a_2=a_3=0$ and $a_1=\pm 1$.  Varying the values of this constant, we can estimate how different initial configurations deviate from our resummed result.

\begin{figure}[t]

  \includegraphics[width=0.5\linewidth]{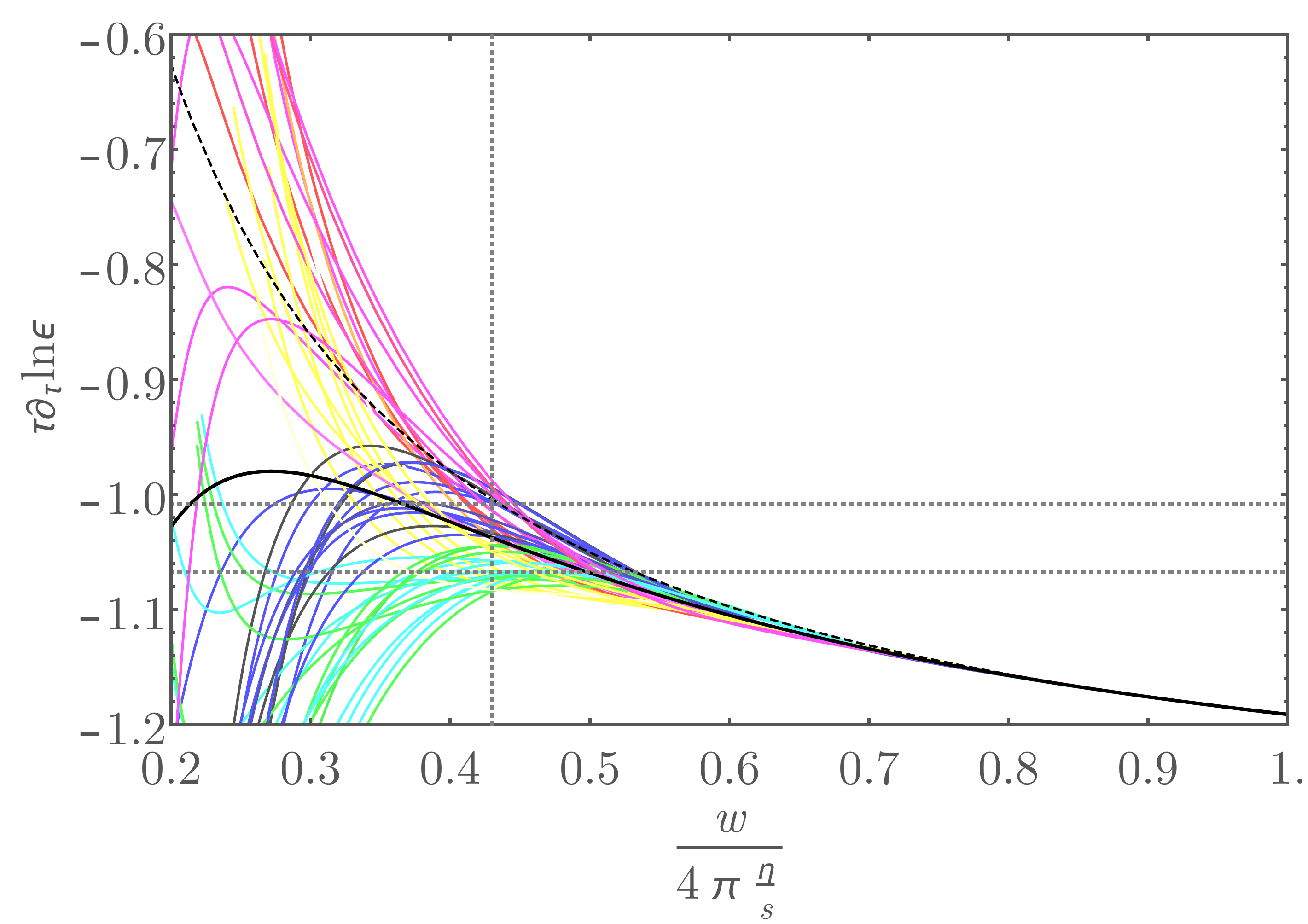}
  \put(-70,115){$\lambda_{GB}=0$}
  \includegraphics[width=0.5\linewidth]{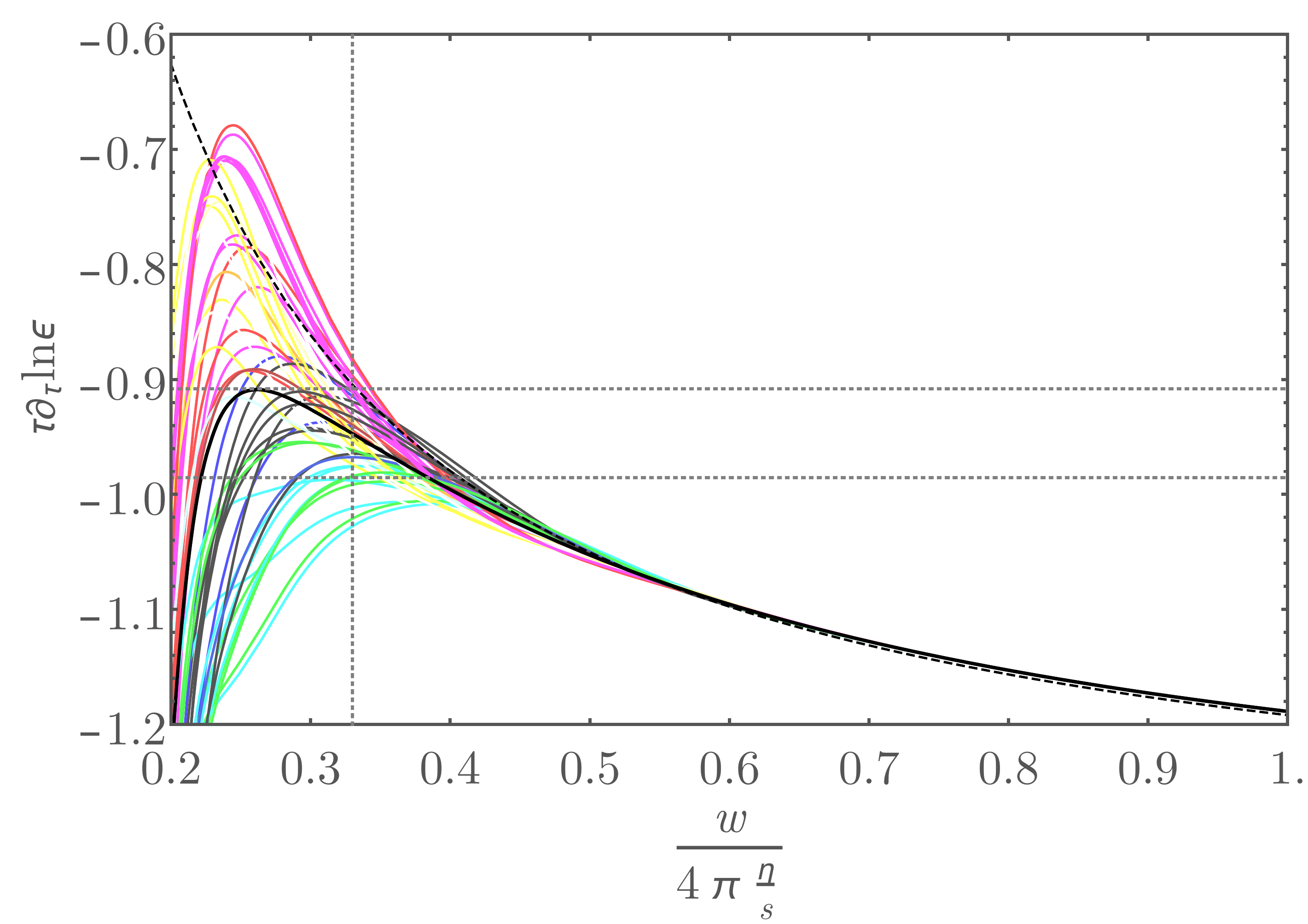}
  \put(-70,115){$\lambda_{GB}=-0.1$} \\
  \includegraphics[width=0.5\linewidth]{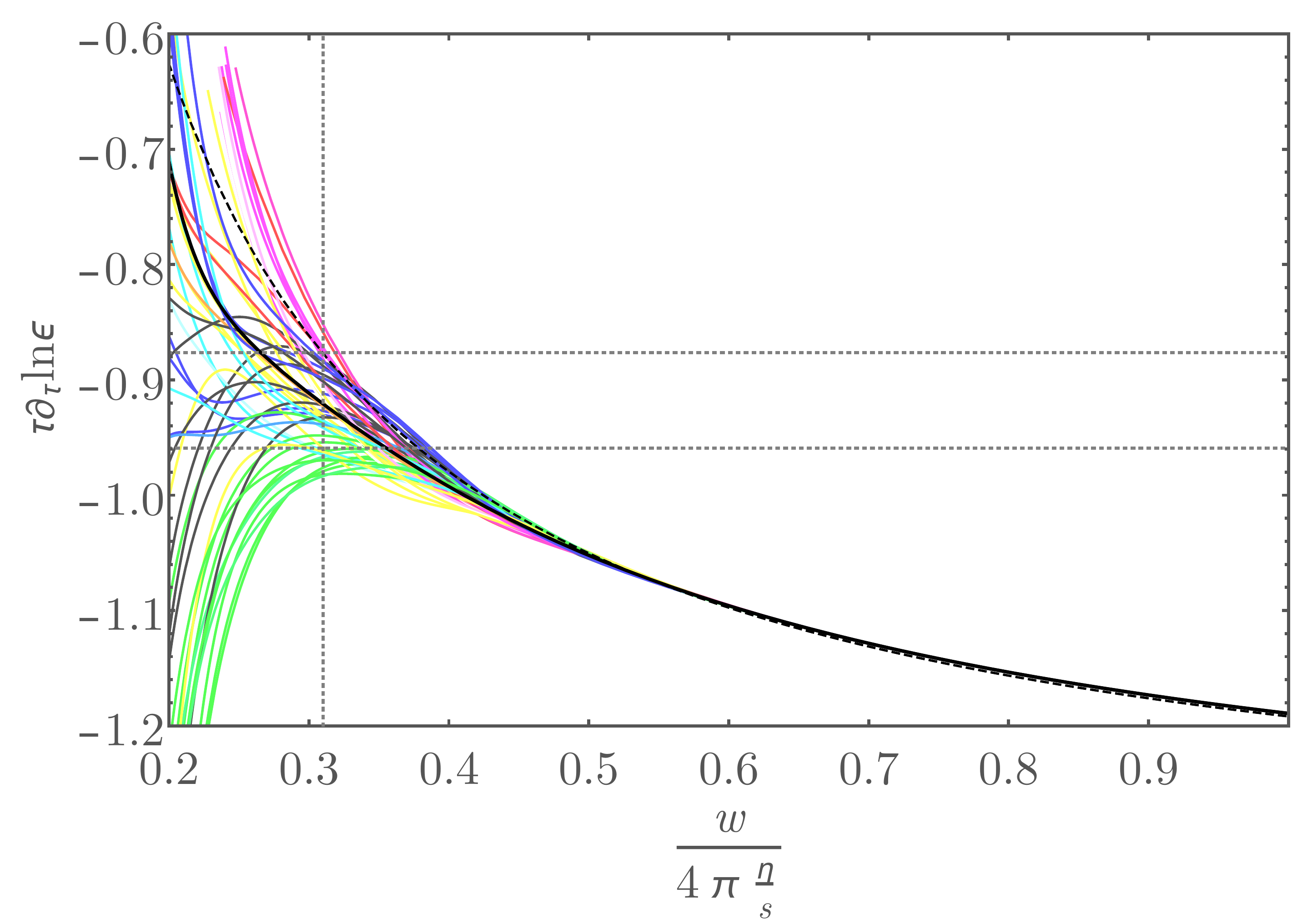}
    \put(-70,115){$\lambda_{GB}=-0.2$} 
  \includegraphics[width=0.5\linewidth]{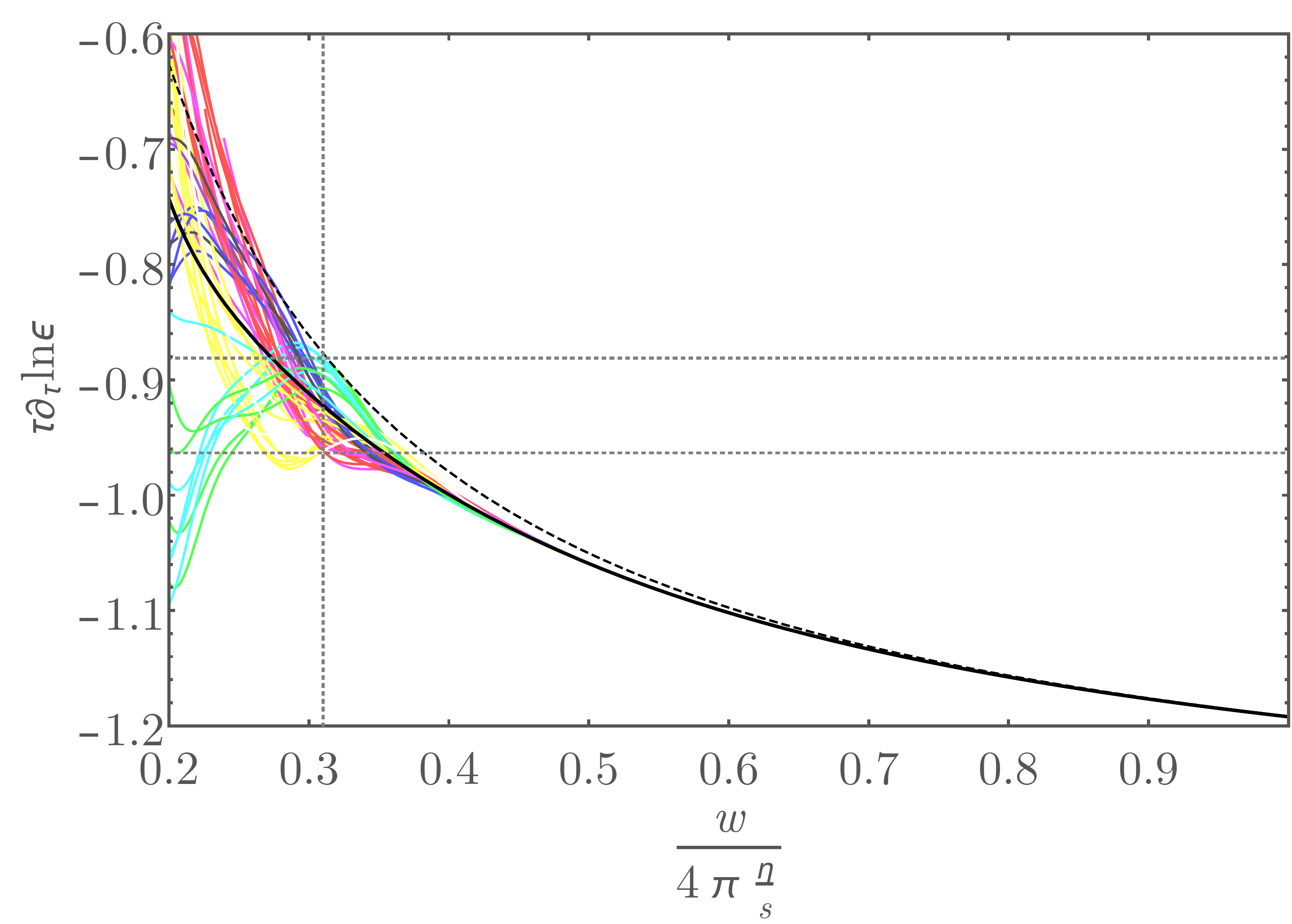}
    \put(-70,115){$\lambda_{GB}=-0.5$} \\
  \includegraphics[width=0.5\linewidth]{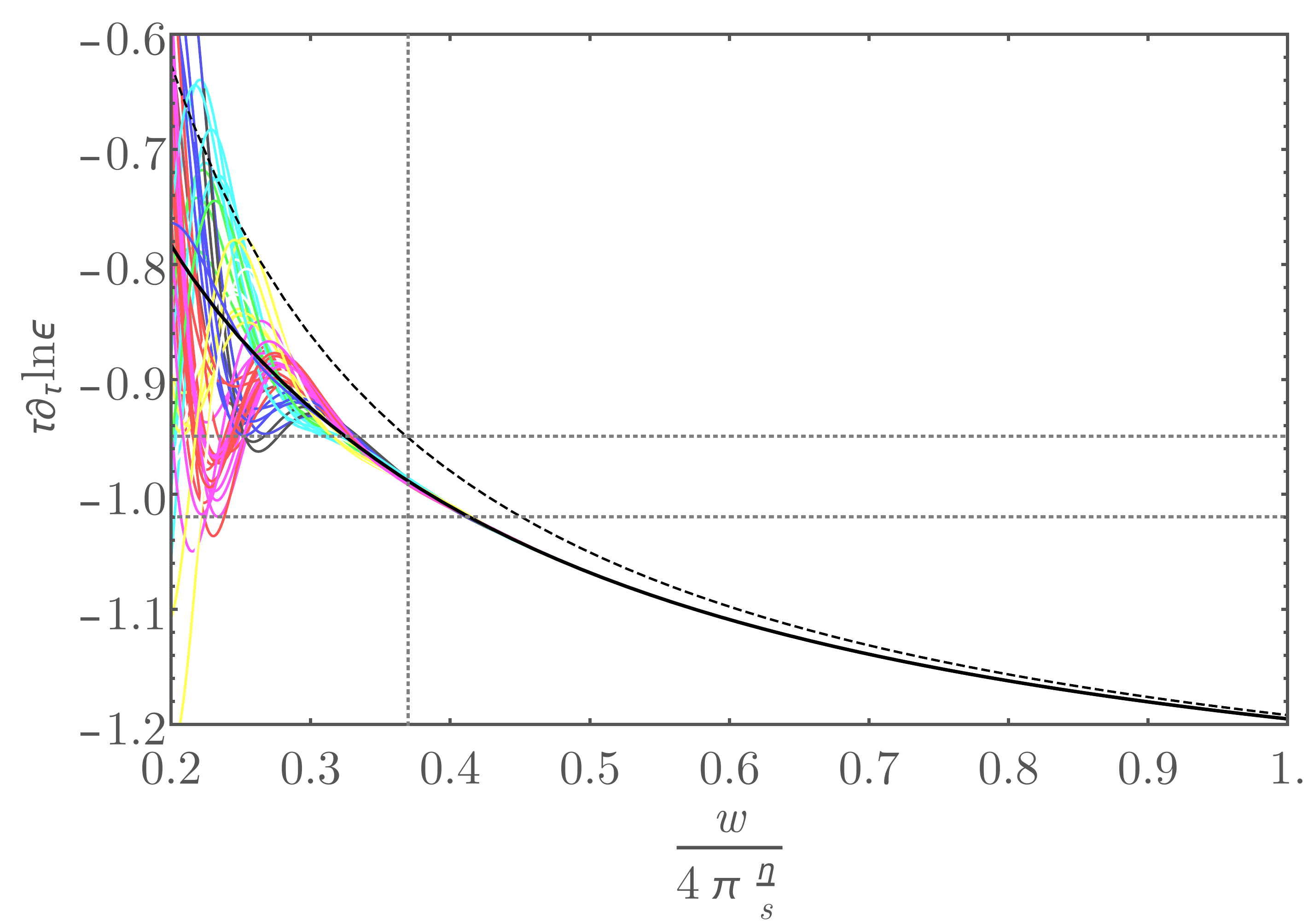}
      \put(-70,115){$\lambda_{GB}=-1$} 
 \includegraphics[width=0.5\linewidth]{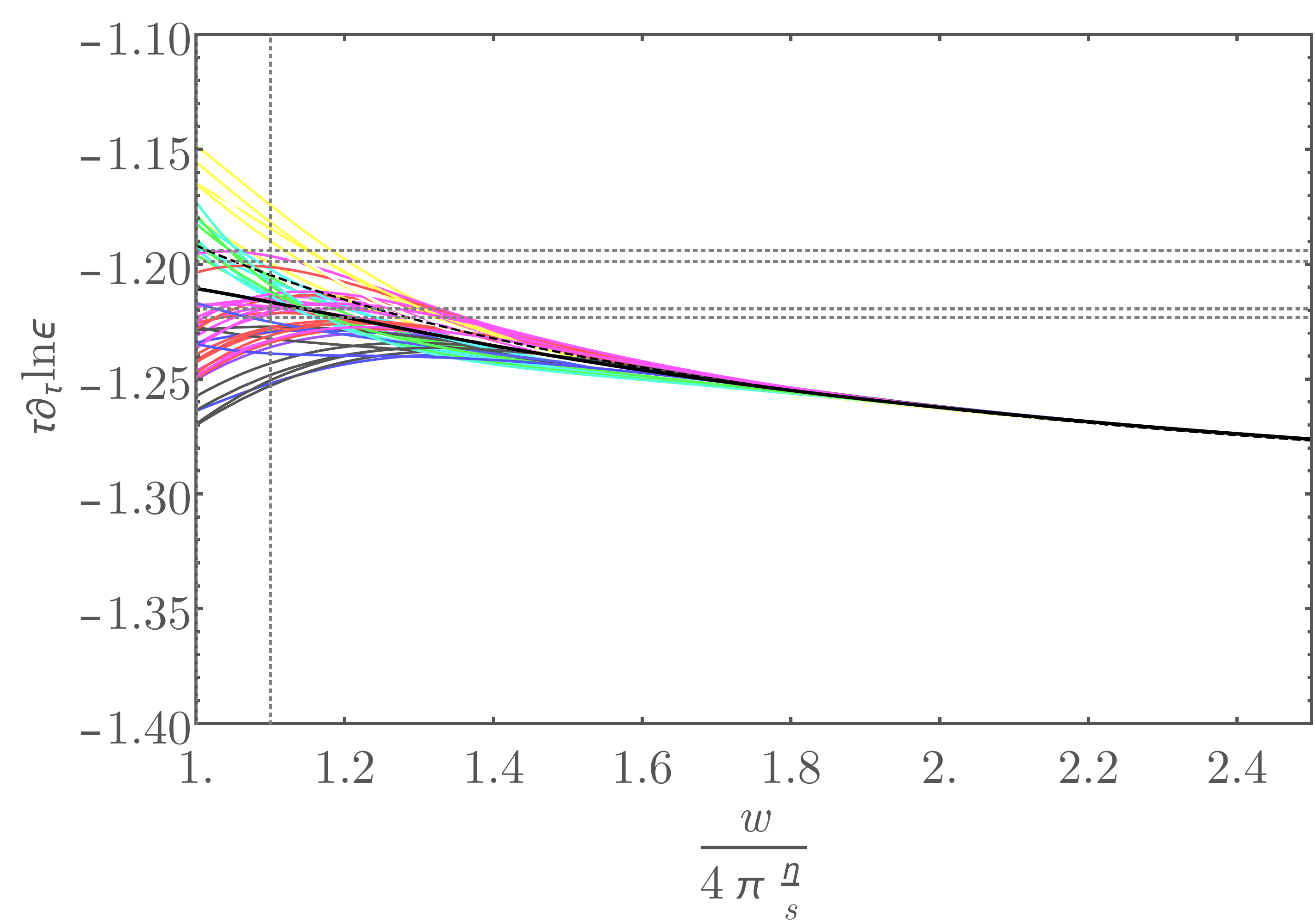}
       \put(-120,115){RTA Kinetic Theory} \\
  \caption{\label{fig:att_estimate}
Logarithmic derivative of the energy density as a function of the viscosity-rescaled inverse gradients for different values of $\lGB$. In all panels, the thick solid lines are the result of the resummation of the hydrodynamic series, while the dashed line correspond to the first order hydrodynamic prediction. The thin coloured lines correspond to adding to the resummation the discontinuities defined in \eqn{eq:disc_def} with arbitrary coefficients. 
The vertical dotted line indicates the rescaled hydrodynamization time $\frac {w_{\rm hyd}s }{4 \pi \eta}$, the horizontal lines give the corresponding $R=(1\pm 0.1)\left . R \right|_{w_{\rm hyd}} $ as displayed in Table (\ref{table}) (including its uncertainties for RTA).
 Every curve corresponding to evolution in a Holographic theory is insensitive up to $2\%$ to the choice of the Pad\'e order $N$ used for $w > 0.3$. For the case of $\N=4$ there is no visible change for the entire region plotted. All RTA curves plotted were insensitive to $N$ at the level of $1\%$.}
\end{figure}

In \fig{fig:att_estimate} we explore the effect of different initial conditions on the time evolution of boost invariant expansion of different theories. The discontinuities discussed before yield a characteristic magnitude of the size of the non-perturbative corrections needed to appropriately define the trans-series. We will vary the coefficients $a_i \in \left(-1,1\right)$ to gauge the spread of typical initial conditions of such time evolution\footnote{We have checked that this procedure leads to a spread in anisotropy paramater comparable to that induced by the different initial conditions in $\N=4$ SYM reported in \cite{Spalinski:2017mel}.}. Following different extractions of attractor solutions in the literature, \cite{Heller:2015dha,Romatschke:2017vte,Romatschke:2017acs}, the attractor may be identified by the so called ``small roll'' condition, which demands that the the time derivative $f'$, with $f$ defined in  \eqn{eq:fdef}, is small all along the evolution of the system. For this reason, in \fig{fig:att_estimate} we show the logarithmic derivative of the energy density, $\tau \partial_{\tau} ln \epsilon$, from which such a derivative may be inferred. In these plots, the solid thick line corresponds to the resummation, while the colourful thin lines correspond to different time evolutions obtained by varying $a_i$.  In all panels, the dashed line corresponds to the first order hydrodynamic prediction for this quantity. Finally, the vertical dotted line marks the hydrodynamization time extracted in Table~(\ref{table}) and the horizontal dotted lines indicate the values of $\tau \partial_{\tau} ln \epsilon$ which correspond to varying $R$ by $10\%$ around the resummation.

The inspection of this figure shows that in all holographic calculations the result of the resumation provides a good proxy to the attractor at hydrodyanamization time. For all values of $\lGB$, the variation within typical initial conditions of the evolution of the energy density is approximately within the hydrodynamization criterium used to determine the $w_{\rm hyd}$. This shows that our extraction of $w_{\rm hyd}$ is trustable; in additions, assuming that the attractor is captured by this set of typical initial conditions, the attractor should also be approximated to better than 10\% after this time. Furthermore, the spread of the different initial conditions also show that, while specific initial conditions may converge faster,  for typical configurations, we only expect convergence towards the attractor when this special solution is well described by first order hydrodynamics. Note also that increasing the range of $a_i$ only make this conclusion stronger; for generic initial conditions the properties of the attractor prior to hydrodynamization do not strongly affect the time evolution of the system. 

We close this section by noticing that the RTA computation exhibits a much stronger dependence on initial conditions that the holographic computations\footnote{
We thank M. Heller, M. Spali{\'n}ski  and V. Svensson for private communication on recent  analysis  of RTA kinetic theory with a non-conformal relaxation time \cite{HellerTBA} which exhibits a trans-series structure with multiple independent contributions with identical exponential suppressions.
This indicates that for conformal RTA the trans-series may be more complicated than what we have assumed in this paper.
}. For RTA not only our estimated hydrodynamization time occurs much later than for the holographic computations, but also at this larger value of the re-scale gradient the spread of initial conditions is large and many of the individual initial conditions are not well approximated by first order hydrodynamics. This implies that the relevance of the attractor for individual initial conditions becomes important at values of the gradient when the attractor is better approximated by hydrodynamics. As explicitly shown in 
Appendix~\ref{sec:npm}, the origin of this large spread is the ambiguity associated to the complex poles in the Borel plane, since these posses much larger residues than the real poles. Remarkably, the origin of these poles is unclear, since they do not appear in a linear analysis \cite{Heller:2016rtz}. 

\section{Discussion}
\label{sec:conclusions}

Understanding the unexpected success of hydrodynamics to describe the off-equilibrium dynamics of interacting systems is an important challenge, not only theoretically but also with important practical applications to heavy ion physics and beyond. 
To address the success of this low energy effective theory much beyond its expected regime of validity, the emergence of special time-dependent configurations of the interacting theory, which act as attractors for all possible system evolutions and which generalise the hydrodynamic expansion beyond the limit of small gradients, has been suggested as a possible explanation. Motivated by this suggestion, in this paper we have applied the extension of hydrodynamics beyond the gradient expansion to the boost invariant flow of the field theory dual of Gauss-Bonnet gravity in 5D, which may be viewed as a laboratory to study finite coupling corrections to infinitely strongly coupled theories. 

As we have already stressed, we have chosen to analyse Gauss-Bonnet holography since, at least in principle, it allows us to explore non-perturbative values of 
$\lGB$ the parameter that controls higher-derivative corrections to Einstein gravity. We would like to remark once again that the holographic dual to this theory is unknown; and it is not even clear whether in that putative dual theory the terms included via non-vanishing $\lGB$ correspond solely to t'Hooft coupling corrections or if they also include finite corrections in the rank of the gauge group $N_c$. Nevertheless, the relaxation dynamics of non-hydrodynamic modes at finite (and negative) $\lGB$ exhibits qualitative similarities to the effect of finite t'Hooft coupling corrections for those dynamics in $\N=4$ SYM. In particular, both theories exhibit purely dissipative relaxation channels which, from the point of view of holography, are due to higher curvature terms. Note also that our analysis has been performed for values of $\lGB$ beyond the causality bounds of \cite{Brigante:2007nu,Buchel:2009tt}. 
For this reason, we are not able to explore the boost invariant expansion in the $\tau \rightarrow 0$ limit.
Nevertheless,  since the unphysical behaviour of Gauss-Bonnet gravity occurs in the ultra-violate, we have concentrated our analysis around how different field configurations approach the hydrodynamic regime.

One of our main results is the analysis of singularities of the Borel transform of the hydrodynamic series in this theory. In accordance with the general theory of resurgence, and as already observed in $\N=4$ SYM, these singularities reflect the characteristic QNM frequencies that control the relaxation of small non-hydrodynamical excitations. 
A direct consequence of the new purely dissipative modes  is the presence of  singularities on the real $\xi $ axes, the variable conjugate to the inverse gradient. These, together with the complex singularities associated to other QNM's of the dual theory make the analytical structure of the fixed coupling calculation richer than in the infinite coupling limit. But even more importantly, the structure of singularities qualitatively interpolates between the infinite coupling limit obtained via holography and the weakly coupled limit, obtained by kinetic theory in the RTA. This may be viewed as an additional motivation to study the large order gradient expansion in this higher-derivative theory. 

To explore the effect of this analytic structure on the early time dynamics of the system, we have resummed the hydrodynamic series via Borel-Pad\'e techniques. 
This allows us to extend the information in the large order gradient expansion to the large gradient region, for values of $w$ such that the contribution of increasing orders in the gradient expansion lead to large, alternating contributions.
Remarkably,  as in all other examples studied in the literature, the resummation of the gradient expansion of the field theory dual to this high-derivative gravity is approximated by first order hydrodynamics at an unexpectedly early time, in a region where viscous effects are large. At this hydrodynamization time, the pressure anisotropy in the expansion is comparable in all strongly coupled computations, independent of $\lGB$, which implies that the hydrodynamization occurs at comparable viscosity-scaled gradients, $w s /\eta$. By comparison, our analysis of the RTA kinetic theory gradient expansion computed in \cite{Heller:2016rtz} indicates that hydrodynamization occurs later, even in the viscosity-scaled gradient, at smaller values of the anisotropy parameter, although these are still large. Our results are consistent with the numerical solutions of RTA 
described in \cite{Heller:2016rtz}.

This resummation allows us to explore the dynamics of the hydrodynamic attractor in this holographic model. 
Certainly,  resummation techniques cannot solely determine the behaviour of the attractor. To fully determine this configuration, analysis able to explore the $w\rightarrow 0$ limit must be performed.  However, at sufficiently late times, when all non-perturbative contributions have relaxed, the resummation must coincide with the attractor. To gauge how close the resummation is from the attractor, we have estimated the relaxation of transient behaviour by studying the discontinuities of the inverse Borel transform over different contours of integration. Since those discontinuities must be cancelled by non-perturbative contributions, these provide a natural scale for the magnitude of these corrections. By varying the magnitude of these modes we can gauge the deviation from the resumation of generic initial conditions. This procedure may be also understood as varying the contribution of the leading QNM over the evolving system. From this analysis we conclude that in all holographic computations, the expected deviation of generic initial conditions from the resummation at hydrodynamization time is comparable to the difference between the resummation and first order hydrodynamics.  As a consequence, our resummation will be a good approximation to the attractor at hydrodynamization time; however, at earlier times this not may be the case\footnote{In fact, our resummation for $\N=4$ SYM differs from the attractor found in \cite{Romatschke:2017vte} at $w<0.4$, as also found by Spali{\'n}ski \cite{Spalinski:2017mel}.}. We may therefore conclude that while individual configurations may converge to the attractor earlier, in all these strongly coupled computations the relaxation of generic initial conditions occurs whenever the system has hydrodynamized. Our analysis also suggests that the sensitivity of kinetic theory to initial conditions persists up to significantly smaller viscosity-rescaled gradients.

Finally, we would like to conclude with an intriguing observation. By analysing the magnitude of the discontinuities in different directions in the complex plane we can estimate the dominant source of initial data dependence in the late time transient behaviour. Surprisingly, for all the values of $\lGB$ studied the dominant contribution is always associated with the complex QNM, which leads to complex singularities in the Borel Plane. The pure dissipative mode is always subleading, even for large negative values of $\lGB$, when the associated singularity is close to the origin and therefore does not possess an obvious suppression (see \eqn{eq:trans}). The numerically extracted discontinuities are shown in Appendix~\ref{sec:npm}. What is even more remakable is that an identical behaviour is observed in kinetic theory, where the dissipative poles are much closer to the origin that the complex ones. This is even more surprising after realising that in RTA it is only the pure dissipative mode that can be obtained from linear response, while the origin of the complex singularities is not yet understood. It would be interesting to explore the effects of this behaviour in other observables. 


\acknowledgments

We thank A. Starinets for suggesting the problem. We thank A. Kurkela for providing us with the kinetic theory coefficients. We also thank M. Heller, R. Janik, P. Romatschke, and M. Spali{\'n}ski for useful discussions. 
JCS is a University Research Fellow of the Royal Society. BM is a Commonwealth Scholar and is also supported by the Oppenheimer Fund Scholarship. N. I. G. was partially supported by the Royal Society research grant "Strange Metals and String Theory" (RG130401) and also by the European Research Council under the European Union's Seventh Framework Programme (ERC Grant agreement 307955).



\appendix
\newpage
\section{Power Series Solution to all Orders} \label{GenSoln}

While constructing solutions for our bulk geometry we found that specific redefinitions of the metric coefficients allowed us to express our equations of motion as $\lambda_{GB}$-independent linear operators sourced by $\lambda_{GB}$-dependent functions. It is easy to see that these linear operators are in fact those of the $\lambda_{GB} = 0$ case which have exact solutions in terms of Greens functions. We could not however find a closed form expression for each source for arbitrary order, and each solution generically has explicit dependence on solutions at all orders below it.

Starting from the ansatz given in Eq. (\ref{eq:genBImetric})
\be
 ds^2 = -r^2 A(\tau,r)d \tau^2 + 2 d r d \tau + (r \tau+1)^2 e^{b(\tau,r)} dy^2 + r^2 e^{ c(\tau,r)} d x_{\perp}^{2}
\ee
we make a further redefinition $d(\tau,r) = c(\tau,r) + \frac{1}{2} b(\tau,r)$ and expand the unknown functions as power series' in $u= \tau^{-2/3}$,   
\begin{align}
A(\tau,r) & = \sum_{i=0} u^{i} A_{i}(s)  , \\
b(\tau,r) & = \sum_{i=0} u^{i} b_{i}(s) ,\\
d(\tau,r) & = \sum_{i=0} u^{i} d_{i}(s) ,
\end{align}
where $s = 1/(r\tau^{1/3})$. We can solve the Field equations of Gauss-Bonnet  perturbatively at each order in $u$, for which we will find 3 e.o.m for $A_{i}(s)$, $d_{i}(s)$ and $b_{i}(s)$, and 2 constraint equations that we will evaluate at $s=1$. The $i=0$ solutions are given by a standard black-brane metric solution stated in Eq.'s (\ref{eq:zerothOrder_1}) to (\ref{eq:zerothOrder_2}).
For each order $i\geq 1$ the functions $A_{i}(s)$, $d_{i}(s)$ and $b_{i}(s)$ must satisfy linear second order ODE's of the form
\begin{align}
\mathcal{L}_{\lambda}^{d}(d_{i}) & = j_{i}^{d} , \\
\mathcal{L}_{\lambda}^{A}(A_{i}) & = j_{i}^{A} , \\ 
\mathcal{L}_{\lambda}^{b}(b_{i}) & = j_{i}^{b} ,
\end{align}
where $\mathcal{L}_{\lambda}^{f}$ is a linear operator depending on $\lambda_{GB}$ which will act on function $f_{i}$ to give the source $j_{i}^{f}$. For all $i \ge 1$ we impose that $A_{i}(s)$, $b_{i}(s)$ and $d_{i}(s)$ all vanish at the boundary ($s = 0$) and are regular at the horizon which we fix (through co-ordinate reparameterization invariance) to be at $s=1$. A consequence of this choice of co-ordinates is that $A_{i}(1)=0$ for $i\geq 1$ so that the constrain equations take the simple forms
\begin{align}
d_{i}(1) & = J_{i} , \\
A_{i}^{'}(1) & = -2 (1-4 \lambda_{GB}) K_{i} ,
\end{align}
where $J_{i}$ and $K_{i}$ are functions of $\lambda_{GB}$, and primes denote derivatives with respect to $s$. Under the redefinitions\footnote{The linear operator $\mathcal{L}^{b}_{\lambda}$ contains only derivatives in $s$ so for convenience we will treat $b^{'}_{i}$ as the function we are solving for.}
\begin{align}
\label{eq:rescale1}
d_{i} & = \tilde{d}_{i} , \\ 
A_{i} & = \frac{1-4\lambda_{GB}}{\sqrt{1-4\lambda_{GB}(1-s^4)}}\tilde{A}_{i} , \\ 
\label{eq:rescale3}
b^{'}_{i} & = \left(\frac{2 \lambda_{GB} \sqrt{1-4\lambda_{GB}(1-s^4)}}{s(1-4\lambda_{GB})(1-\sqrt{1-4\lambda_{GB}(1-s^4)})}\right) \tilde{b}^{'}_{i} , 
\end{align}
the equations of motion become
\begin{align}
\label{eq:lambda0L1}
\mathcal{L}_{0}^{d}(\tilde{d}_{i}) & = \tilde{j}_{i}^{d} , \\
\mathcal{L}_{0}^{A}(\tilde{A}_{i}) & = \tilde{j}_{i}^{A} , \\ 
\label{eq:lambda0L3}
\mathcal{L}_{0}^{b}(\tilde{b}^{'}_{i}) & = \tilde{j}_{i}^{b} , 
\end{align}
where the linear operators $\mathcal{L}_{0}^{C}$ no longer have a dependence on $\lambda_{GB}$ and the source terms $j_{i}^{C}$ are scaled as
\begin{align}
\tilde{j}_{i}^{d} & = \frac{1}{\sqrt{1-4\lambda_{GB}(1-s^4)}} j_{i}^{d} , \\
\tilde{j}_{i}^{A} & = \frac{1}{1-4\lambda_{GB}} j_{i}^{A} , \\ 
\tilde{j}_{i}^{b} & = j_{i}^{b}.
\end{align}
The $\lambda_{GB}$-independent linear operators are
\begin{align}
\mathcal{L}_{0}^{d} & = \partial_{s}^{2}\\
\mathcal{L}_{0}^{A} & = s^2\partial_{s}^{2}-5 s\partial_{s} + 8 \\ 
\mathcal{L}_{0}^{b} & = \frac{1}{4}s\partial_{s} - 1.
\end{align}
Equations (\ref{eq:lambda0L1}) to (\ref{eq:lambda0L3}) have solutions in terms of Greens functions. 
\begin{align}
\label{eq:perturbativeSoln_d}
\tilde{d}_{i}(s)&= J_{i} s + \left(\int_{0}^{s} dx\, - s \int_{0}^{1} dx\, \right) \int_{1}^{x} dy\, \tilde{j}^{d}_{i}(y) \\
\tilde{A}_{i}(s)&= K_{i} s^2(1-s^2)+\frac{1}{2}s^2 \int_{1}^{s} d x\, \left(\frac{s^2-x^2}{x^5}\right) \tilde{j}^{A}_{i}(x)\\
\label{eq:perturbativeSoln_b}
\tilde{b}^{'}_{i}(s)&=  s^{4}\int_{1}^{s} d x\, \frac{4}{x^{5}} \tilde{j}^{b}_{i}(x)
\end{align}
Here integration constant associated with $\tilde{b}'_{i}$ has necessarily been fixed to $0$ to ensure regularity of $b'_{i}$ at the horizon.

One can retrieve all perturbative solutions iteratively by finding $\tilde{d}_{i}$, $\tilde{A}_{i}$ then $\tilde{b}^{'}_{i}$ and substituting the results to find $d_{i}$, $A_{i}$ and $b^{'}_{i}$.\footnote{One can find $b_{i}$ through a an integral setting the integration constant such that $b_{i}(s=0) = 0$.}

The first few sources are given by
\begin{align}
\tilde{j}^{d}_{1}(x)& = 0 , \\
\tilde{j}^{A}_{1}(x)& = -2 x ,\\
\tilde{j}^{b}_{1}(x)& = \frac{1}{4}x\left(-14+\frac{3}{\lambda_{GB}}+8\lambda_{GB}\right)+\frac{x(1-4\lambda_{GB})(1-2 \lambda_{GB})(-3 + 4 \lambda_{GB} (3- 5 x^4))}{4\lambda_{GB}(1-4\lambda_{GB}(1-x^4))^{3/2}} ,
\end{align}
with
\begin{align}
J_{1} & = -(1-2\lambda_{GB}) , \\
K_{1} & = -1 .
\end{align}

Solving Eq.'s (\ref{eq:perturbativeSoln_d}) to (\ref{eq:perturbativeSoln_b}) at order $i=1$ and then rescaling the results using Eq.'s (\ref{eq:rescale1}) to (\ref{eq:rescale3}) we can recover
\begin{align*}
d_{1}(s) & = -(1-2\lambda_{GB})s , \\
A_{1}(s) & = -\frac{2}{3} \frac{s(1-s^3)(1-4\lambda_{GB})}{\sqrt{1-4\lambda_{GB}(1-s^4)}} , \\ 
b_{1}^{'}(s) & = \left(-2(1-2\lambda_{GB})+\frac{2}{3}\frac{4\lambda_{GB} (1-s^3)\sqrt{1-4\lambda_{GB} (1-s^4)}}{1-\sqrt{1-4\lambda_{GB}(1-s^4)}}\right) .
\end{align*}

\newpage

\section{Non-perturbative Modes}
\label{sec:npm}

\begin{figure}[t] 

  \includegraphics[width=0.5\linewidth]{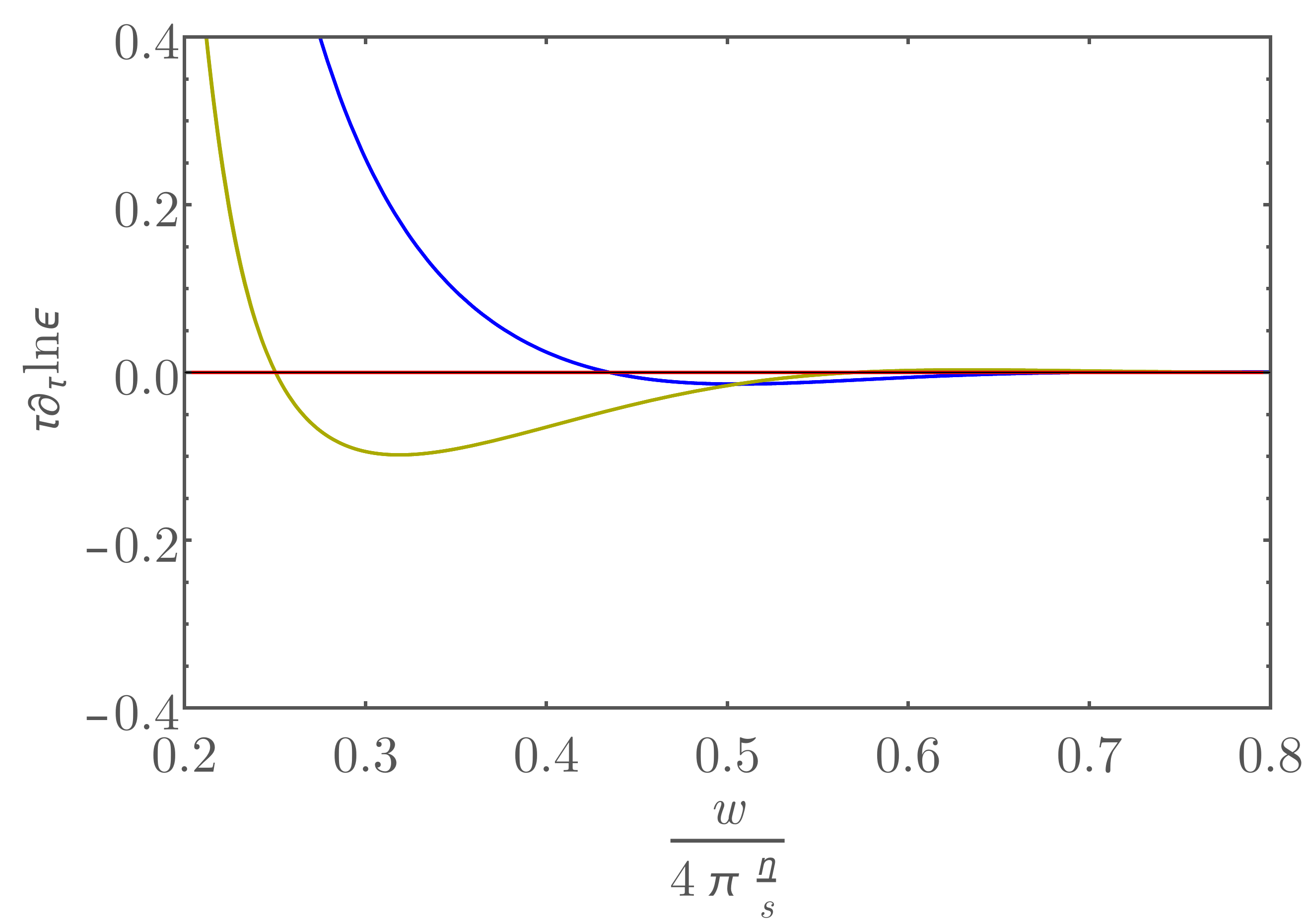}
  \put(-80,115){$\lambda_{GB}=0$}
  \includegraphics[width=0.5\linewidth]{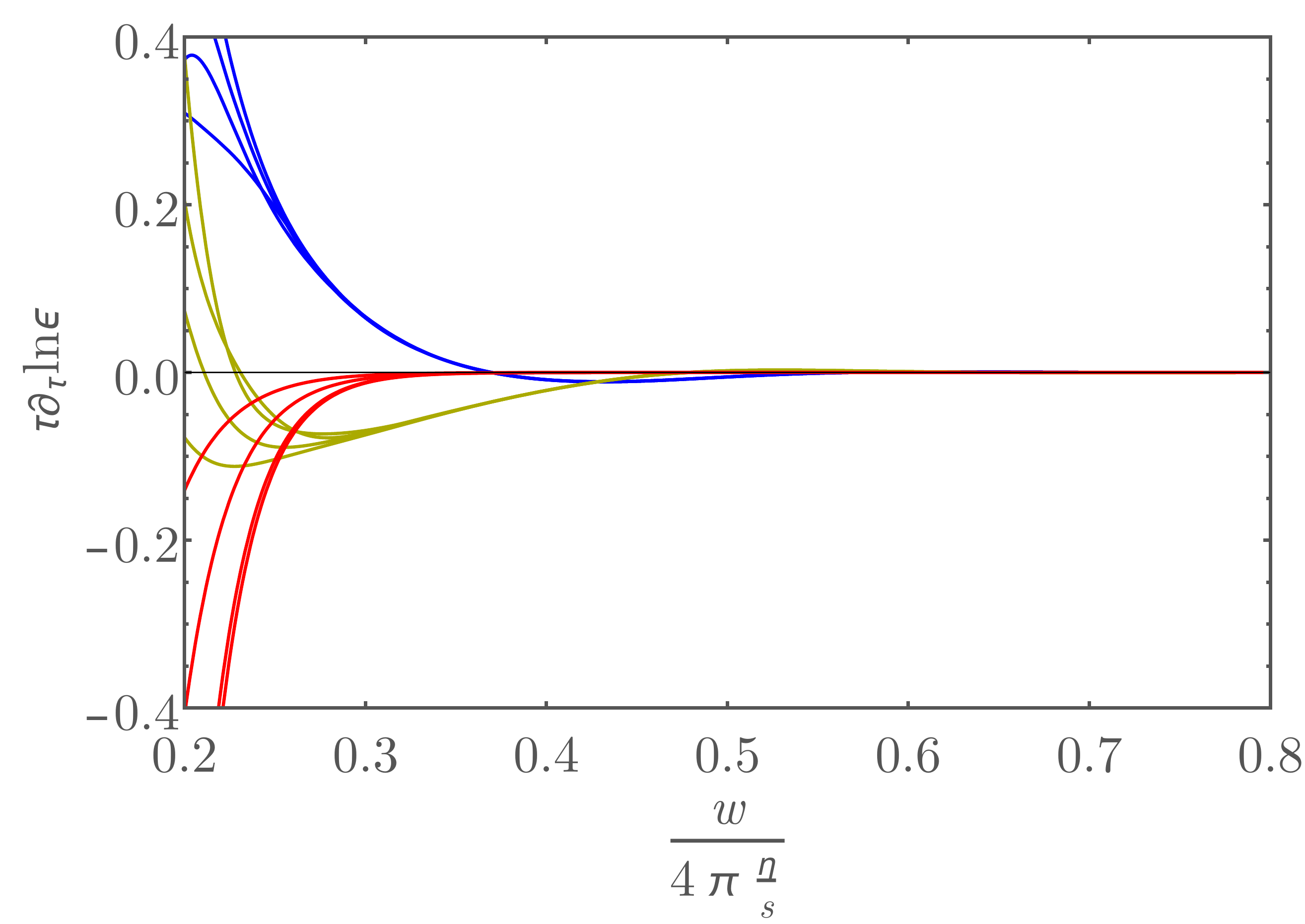}
  \put(-80,115){$\lambda_{GB}=-0.1$} \\
  \includegraphics[width=0.5\linewidth]{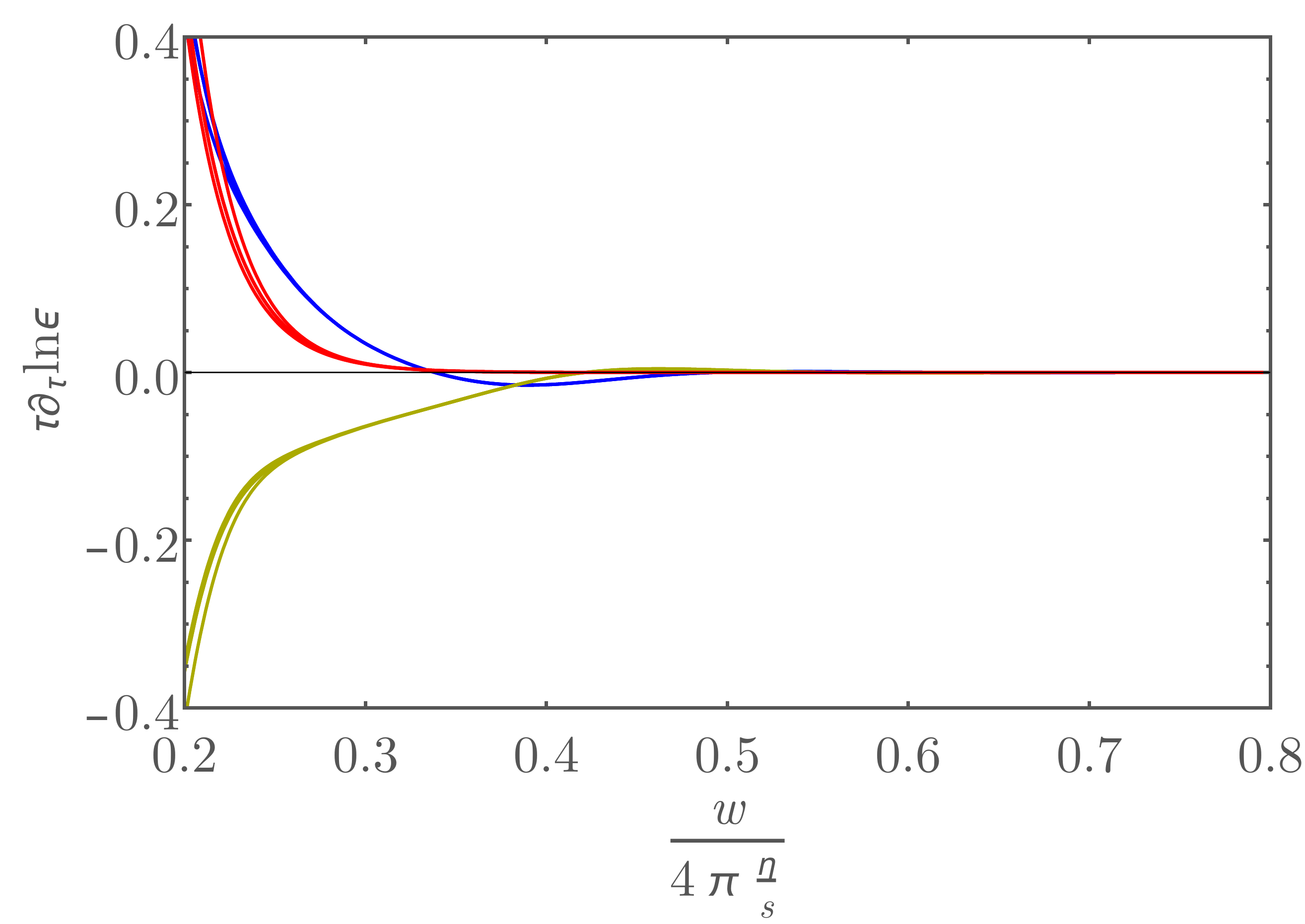}
    \put(-80,115){$\lambda_{GB}=-0.2$} 
  \includegraphics[width=0.5\linewidth]{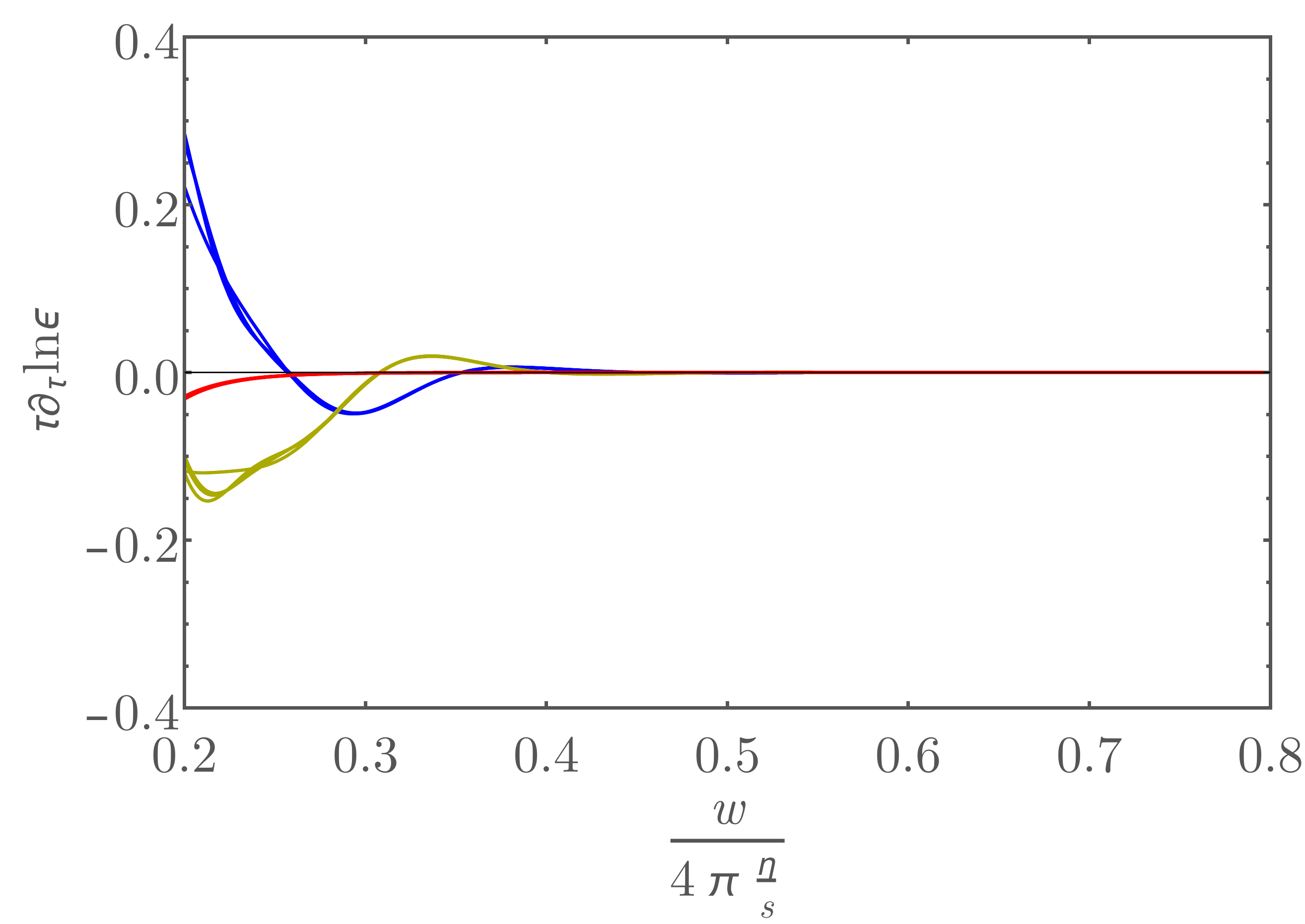}
    \put(-80,115){$\lambda_{GB}=-0.5$} \\
  \includegraphics[width=0.5\linewidth]{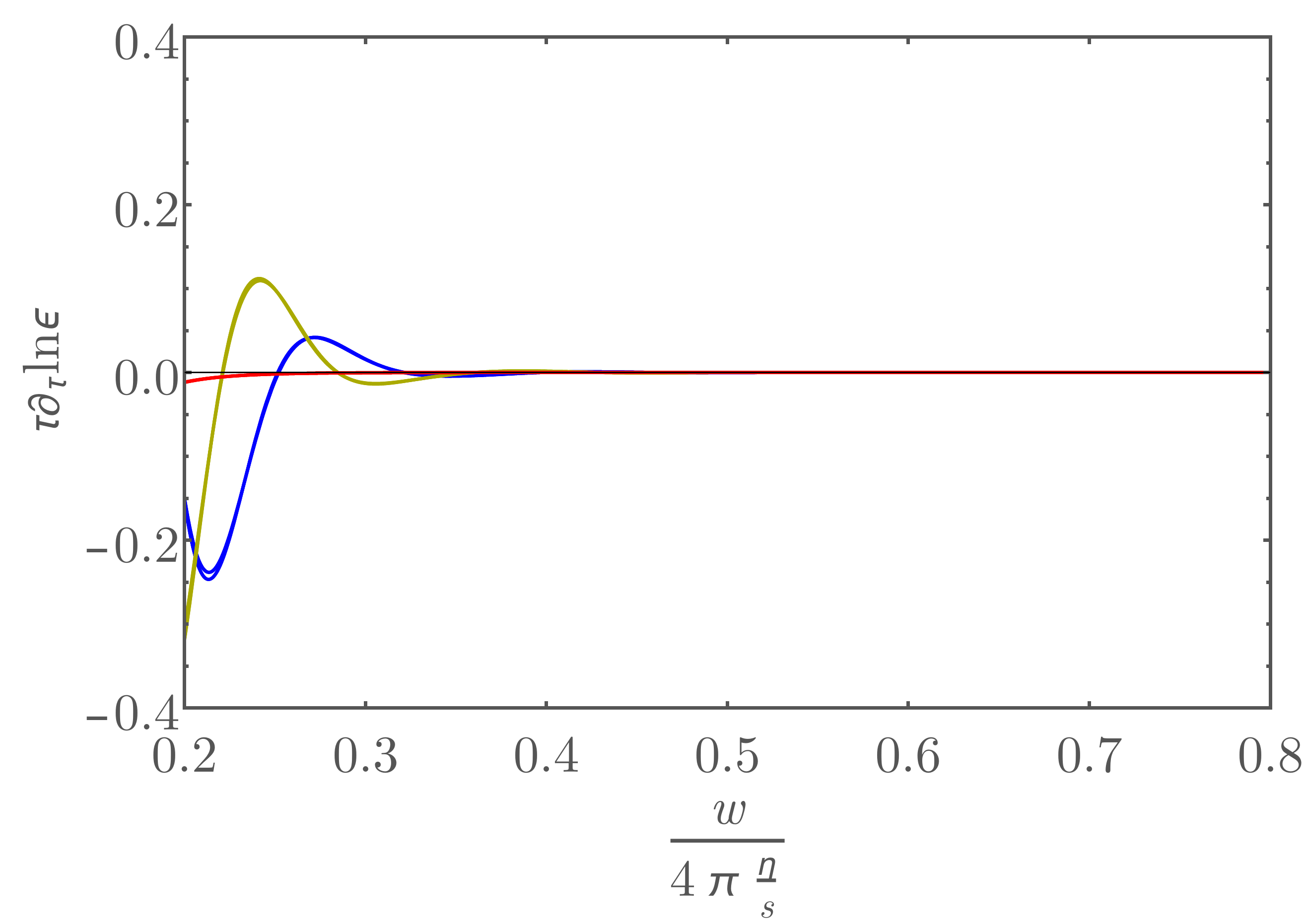}
      \put(-80,115){$\lambda_{GB}=-1$} 
 \includegraphics[width=0.5\linewidth]{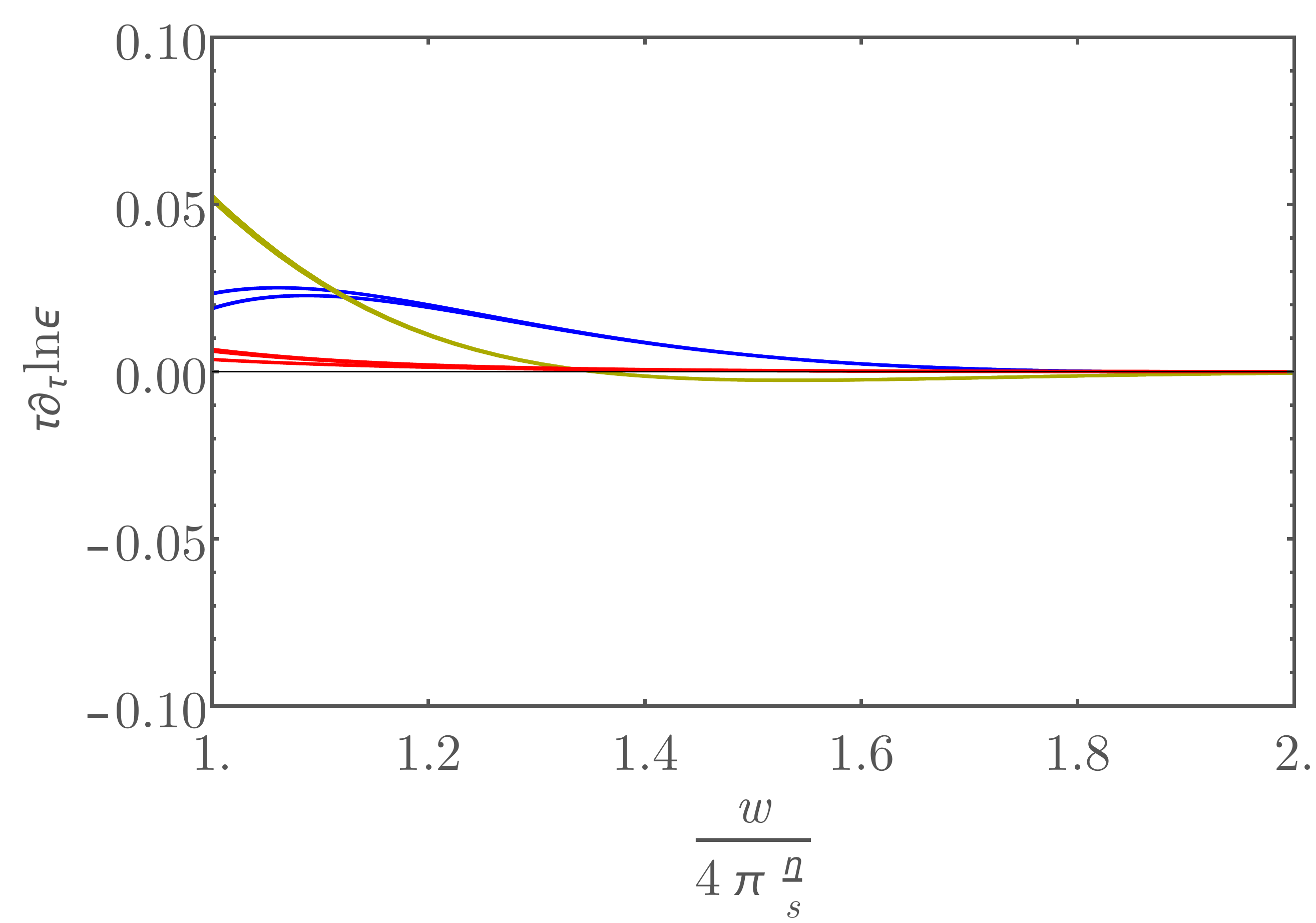}
       \put(-140,115){RTA Kinetic Theory} \\
  \caption{ \label{fig:npms}
  Non-hydrodynamic modes Re[$D_{c}$] (blue), Im[$D_{c}$] (yellow) and $D_{\pm}$ (red) as described in Equation (\ref{eq:disc_def}) for different couplings. Each curve in Fig. (\ref{fig:npms}) has been made by varying the Pad\'e order $N$ over the four previous values, with the exception of the $\lambda_{GB} = 0$ case where $N$ is chosen to take values of $80$, $90$, $120$ and $190$. The visible deviations are intended to give a sense of the convergence of the Pad\'e Approximant and so the convergence of these non-hydrodynamic modes.}
\end{figure}

The discontinuity Re[$D_{c}$], Im[$D_{c}$] and $D_{\pm}$  defined in Equation (\ref{eq:disc_def}) are shown in Fig. (\ref{fig:npms}) in blue, yellow and red respectively for all the models considered. For every case, we use  several values of the Pad\'e order to estimate the uncertainty of this integration.
To avoid ambiguities associated with the oscillatory character of these functions, we estimate the error of each curve by computing the maximum value for different choices of the Pad\'e order $N'=N-\Delta$ with $N=N_{\rm coefficients}/2$ and $\Delta=1,2,3$  of the function
\be
{\rm Err}(w) =\frac{\int^\infty_w dx \left| f_{N}(x)-f_{N'}(x)\right| }{\int^\infty_w dx \left| f_{N}(x)\right|} \,.
\ee
 Using this procedure, we find that dispersive modes, Re[$D_{c}$] (blue) and Im[$D_{c}$] (yellow), deviate for no more than 5$\%$ 
for  $\frac{w}{4\pi \frac{\eta}{s}}> 0.2$, $0.3$, $0.25$, $0.25$, $0.2$, $1.05$ for $\lGB=0$,  $-0.1$, $-0.2$,  $-0.5$, $-1$ and RTA kinetic theory respectively. For the dissipative mode $D_{\pm}$ the curves are accurate to 5\% for $\frac{w}{4\pi \frac{\eta}{s}}> 0.3$, $0.25$, $0.2$ for $\lGB=-0.2$, $-0.5$ and $-1$ respectively. For $\lGB=-0.1$ and for RTA, convergence at 
5\% level is only achieved at $\frac{w}{4\pi \frac{\eta}{s}}> 0.85$ and $2$ respectively. Even though these uncertainties are large, since the dissipative mode is much smaller than the dispersive contribution, these errors do not alter the spread of initial conditions displayed in Fig. (\ref{fig:att_estimate}).

Remarkably, we find that in all cases the dissipative mode is suppressed by at least an order of magnitude relative to the dispersive modes, even when the poles along the real axis in the Borel plane are closer to the origin than the leading complex mode. This is surprising since inspection of the trans-series \eqn{eq:trans} suggests that the contribution of each of the non-perturbative modes to this discontinuity is controlled by the exponential suppression associated to the position of the corresponding singularity in the complex Borel Plane \eqn{eq:Omegadef}. This observation is even more striking for RTA, since the dissipative branch cut in this case is much closer to $\xi=0$ than the complex one, and hence one would expect a larger suppression of those non-linear modes. This suggests that it will be necessity to understand the role of the complex modes in RTA kinetic theory to properly describe the evolution of the system at intermediate to late times.

\newpage

\bibliography{RGBbib}{}
\bibliographystyle{JHEP-2}
\end{document}